\documentclass[11pt]{article}
\usepackage[margin=1in]{geometry}
\usepackage{amsmath,amssymb}
\usepackage{graphicx}
\usepackage{booktabs}
\usepackage{array}
\usepackage{xcolor}
\usepackage{hyperref}
\usepackage{caption}
\usepackage{subcaption}
\usepackage{natbib}
\usepackage{microtype}
\usepackage{enumitem}
\usepackage{float}
\usepackage{placeins}
\usepackage{longtable}
\usepackage{fancyhdr}

\setlist{nosep}
\hypersetup{colorlinks=true,linkcolor=blue!45!black,citecolor=blue!45!black,urlcolor=blue!45!black}
\title{\textbf{Analog Electronic Computation of Stellar Structure:}\\
Polytropic Benchmarks, Circuit Scaling, and Error-Controlled Verification}

\usepackage{authblk}

\author[1,2,*]{Zuhoor Elahi}
\author[1]{Wafa Gull}
\author[1]{Christopher Sirola}

\affil[1]{Department of Physics and Astronomy, University of Southern Mississippi, Hattiesburg, MS, USA}
\affil[2]{Department of Physics, University of Karachi, Karachi, Pakistan}
\affil[*]{Corresponding author: zuhoor.elahi@usm.edu}

\date{}

\begin{document}
\maketitle

\begin{abstract}
We develop an electronic analog-computing formulation for static stellar-structure equations and test it against polytropic Lane--Emden benchmarks with known or high-accuracy reference solutions. Beginning from an exploratory op-amp/multiplier circuit representation of the stellar-structure equations, we identify sign, coefficient, central-regularity, and signal-conditioning issues that prevent a direct hardware interpretation. We then formulate dimensionless benchmark systems, compare alternative first-order state choices, derive regular finite-radius start conditions, and construct voltage-scaled analog architectures subject to finite signal headroom. The principal benchmark family uses the positive derivative state $w=-d\theta/d\xi$ and reproduces the $n=0$, $n=1$, and $n=3$ polytropic surface location $\xi_1$ and dimensionless mass coefficient $\mu_1=-\xi_1^2\theta'(\xi_1)$ at the ideal scaled-voltage level. A common hardware scaling is identified for all three benchmarks, and the nonlinear $n=3$ case illustrates coefficient-preserving headroom repair through reference and computation-speed changes. Local sensitivities and Monte Carlo studies quantify how effective calibration errors propagate to stellar observables and motivate installed-circuit calibration rather than reliance on nominal component tolerances. In a separate regularized $n=1$ device-level bridge, a calibrated regularized-$v$ realization using AD734/OPAx197 vendor macromodels passes all 67 declared deterministic certification cases and three reproducibly seeded 1000-case Monte Carlo populations under the predefined engineering criterion $|\xi_1-\pi|\le0.020$, with zero saturation or convergence failures. The stricter 0.1\% first-zero criterion remains characterization rather than the certification gate. A dimensionless four-state radiative architecture is also demonstrated as an extension beyond polytropes, but is not interpreted as a calibrated real star. Physical hardware validation and vendor execution of the preferred common $w$-form architecture remain separate evidence levels.
\end{abstract}


\section{Introduction}

The equations of stellar structure form a coupled nonlinear differential system in which hydrostatic support, mass conservation, energy generation, and energy transport are solved simultaneously from the center toward the stellar surface. Classical polytropes and the Lane--Emden equation provide a particularly useful reduced setting because they preserve the central mathematical structure of self-gravitating stellar models while supplying exact or accurately tabulated reference solutions \citep{Chandrasekhar1939,Horedt2004}.

Electronic analog computers solve differential equations by mapping state variables to continuously varying voltages and constructing derivatives from integrators, summers, multipliers, dividers, and function generators. Historical electronic analog-computing practice emphasized explicit machine variables, scale factors, initial conditions, and static checks \citep{Bryant1960}. Although digital computation displaced general-purpose analog machines, continuous and hybrid analog computation continues to attract interest in both electronic and wave-based forms \citep{MacLennan2014,UyBui2023}. Historical accounts emphasize that general-purpose electronic analogue computers were widely used as differential-equation and simulation machines in the postwar period, with explicit attention to scaling, interconnection, initial conditions, and error control \citep{Thomason1961,Small1993,Small2001}. More recent work has revisited continuous analog computation using modern integrated electronics and other physical platforms \citep{Tsividis2018,MacLennan2014,UyBui2023}.

The novelty position requires a narrower historical distinction. Analogue computation did appear in astrophysical work: Lema\^itre and Vallarta used Bush's mechanical differential analyzer to integrate cosmic-ray trajectories in the geomagnetic field \citep{LemaitreVallarta1936}. Svein Rosseland then brought differential-analyzer computing directly into an astrophysical institute: the Oslo Analyzer was constructed at the University of Oslo's Institute of Theoretical Astrophysics as a large mechanical differential-equation resource, and its history and scientific role are documented by Holst \citep{Holst1996}. Kopal explicitly acknowledged Rosseland's pioneer work when introducing his 1949 treatment of nonradial oscillations of the standard $n=3$ polytropic model, whose oscillation equations were integrated mechanically on the Rockefeller differential analyzer at MIT \citep{Kopal1949}. These precedents are directly relevant to the history of analogue stellar computation, but the retrieved records do not establish the same component-level \emph{electronic} realization of the static Lane--Emden benchmark family developed here. General electronic differential analyzers likewise became established differential-equation machines in the postwar period \citep{HoweHaneman1953,Nagaraja1957}. In parallel, stellar evolution and stellar-structure calculations rapidly moved onto digital computers and numerical boundary-value methods \citep{Henyey1959,Hoyle1959,LarsonDemarque1964}. A targeted search across these literatures, modern Lane--Emden methods, astronomy/space archives, and a large academic-paper index has not yet identified a directly comparable component-level electronic analogue realization of the \emph{static} stellar-structure/Lane--Emden benchmark family developed here. This is treated only as a bounded negative-search result, not as proof of priority.

The present work asks a narrower question: how can stellar-structure equations be transformed into an electronic analog-computing architecture that remains mathematically transparent, voltage-scaled, quantitatively verifiable, and accompanied by an explicit error budget? The goal is not to replace modern stellar-evolution software. Instead, the project uses stellar physics as a demanding nonlinear benchmark for continuous analog computation.

The development proceeded from an initial Multisim realization of the stellar equations to a sequence of audited equation blocks, exact and numerical benchmark systems, conditioned state-variable formulations, hardware-oriented scaling, nonlinear source construction, and uncertainty propagation. This paper focuses on the static stellar-structure branch. A separate future program addresses stellar pulsation only after the static analog framework is adequately certified.

\section{From stellar-structure equations to an analog architecture}

\subsection{Exploratory circuit concept}

The exploratory circuit treated radius as the independent progression variable and attempted to encode the standard stellar equations directly using operational amplifiers and analog multipliers/dividers. The equations motivating that design included
\begin{align}
\frac{dP}{dr} &= -\frac{Gm\rho}{r^2},\\
\frac{dm}{dr} &= 4\pi r^2\rho,\\
\frac{dL}{dr} &= 4\pi r^2\rho\epsilon,\\
\frac{dT}{dr} &= -\frac{3\kappa\rho L}{64\pi\sigma r^2T^3},
\end{align}
with a separate convective-transport branch in the original schematic.

The initial exercise was scientifically useful because it exposed the difference between an equation drawn as electronic blocks and a reproducible analog-computing realization. Table~\ref{tab:p0audit} summarizes the principal audit results used in the subsequent redesign.

\begin{table}[H]
\centering
\caption{Selected exploratory prototype audit results. The quantitative manuscript benchmarks use the corrected and nondimensionalized descendants rather than the exploratory circuit directly.}
\label{tab:p0audit}
\begin{tabular}{p{0.20\textwidth}p{0.34\textwidth}p{0.35\textwidth}}
\toprule
Equation block & Audit result & Consequence \\
\midrule
Hydrostatic equilibrium & The nonlinear divider topology can represent the $m\rho/r^2$ factor, but the final inverting integration changes the pressure derivative sign under the original voltage convention. & Add an equivalent sign correction or redefine the state polarity. \\
Mass continuity & The two-inversion topology gives the positive derivative, but a unity displayed resistor ratio omits the required $4\pi$ coefficient. & Encode the full coefficient explicitly rather than relying on ambiguous schematic labels. \\
Energy generation & Same $4\pi$ coefficient issue as mass continuity. & Use explicit calibrated gain. \\
Radiative transport & The nonlinear denominator structure is compatible with the target equation, but the final integration polarity reverses the required temperature gradient. & Correct the sign before integration. \\
Stellar center & Direct use of $r=0$ produces electronically singular divider inputs. & Nondimensionalize and begin from a regular finite-radius series state. \\
\bottomrule
\end{tabular}
\end{table}

\begin{figure}[H]
\centering
\includegraphics[width=0.82\textwidth]{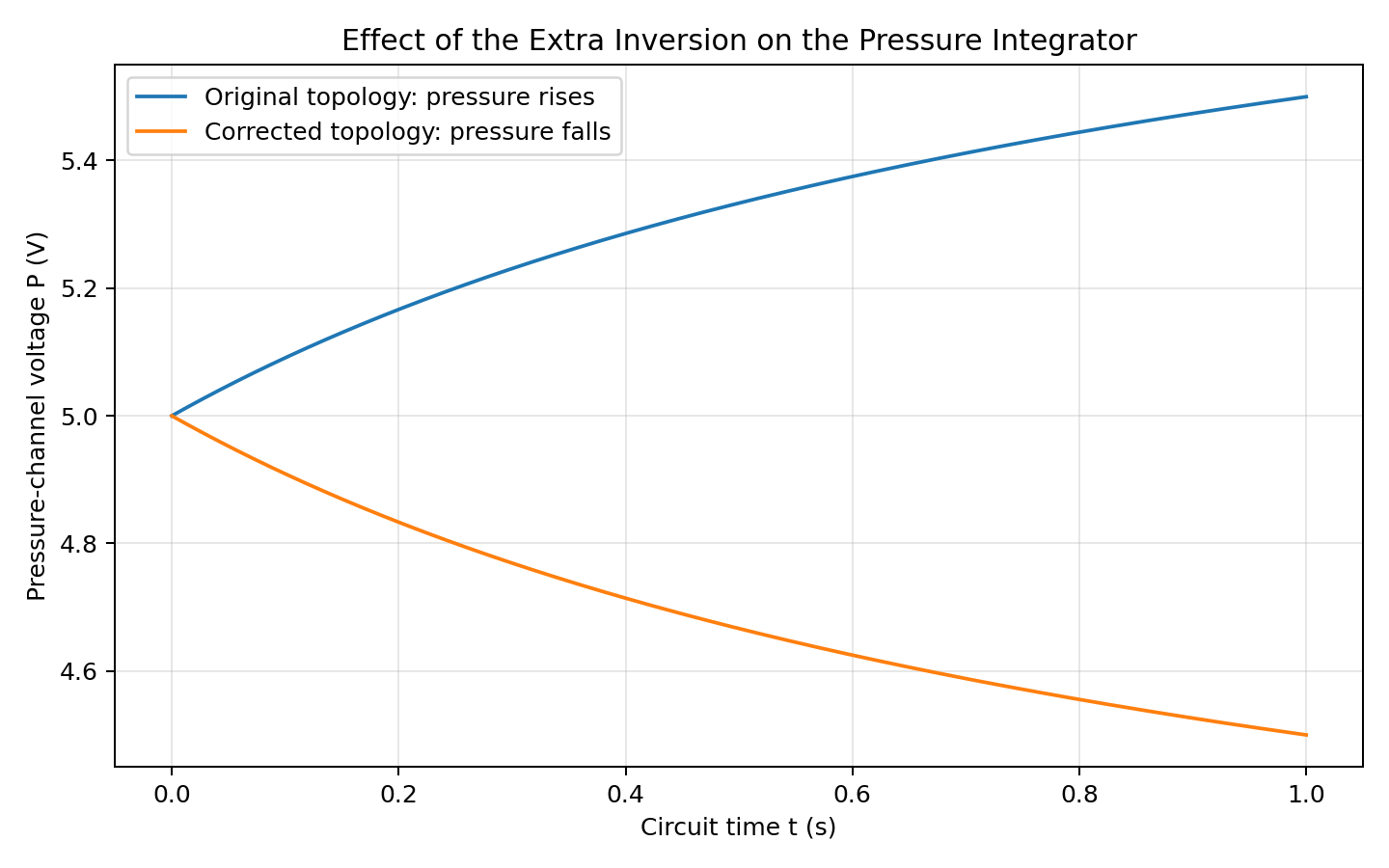}
\caption{Exploratory prototype hydrostatic sign audit. A mathematically valid nonlinear divider can still yield the wrong physical derivative if the integrator polarity is not included in the equation-to-circuit mapping.}
\label{fig:p0sign}
\end{figure}

A simple hydrostatic test already distinguishes the corrected and original sign conventions. For constant normalized $G=m=\rho=1$, $r=1+\tau$, and $P(0)=5$, the corrected solution decreases as $P=4+1/r$, whereas the original polarity increases. Figure~\ref{fig:p0sign} shows this diagnostic.

\begin{figure}[H]
\centering
\includegraphics[width=0.98\textwidth]{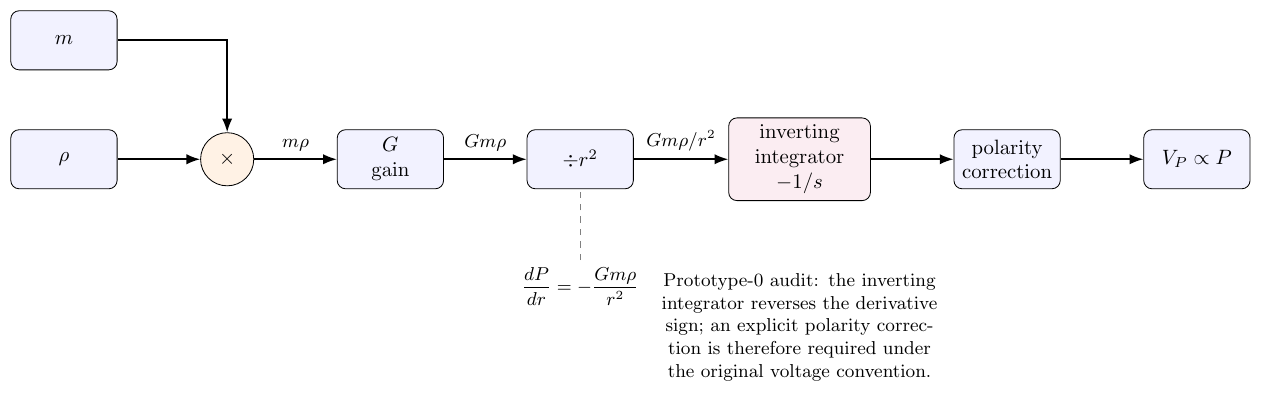}
\caption{Cleaned signal-flow reconstruction of the exploratory prototype hydrostatic-equilibrium branch. The original exploratory Multisim sheets are reproduced in Supplementary Section S1.}
\label{fig:p0clean}
\end{figure}

Figure~\ref{fig:p0clean} restates the hydrostatic branch as an explicit signal-flow circuit map. The purpose of this cleaned diagram is not to replace the archived Multisim sheet, but to make the sign audit reproducible in the paper itself: the product/division chain and the integrator polarity are shown as separate operations.

\section{Analog-computing formulation}

A dimensionless state $q(\xi)$ is represented by
\begin{equation}
V_q=S_q q,
\end{equation}
where $S_q$ is a voltage scale chosen for signal-to-error ratio and headroom. Laboratory time $\tau$ is mapped to the stellar independent variable by
\begin{equation}
\xi=\xi_0+s_\xi\tau,
\end{equation}
so that
\begin{equation}
\frac{dq}{d\tau}=s_\xi\frac{dq}{d\xi}.
\end{equation}
Thus computation speed is an electronics design parameter: it changes the required circuit coefficients without changing the target stellar solution.

The project separates four operations that are often conflated in informal circuit sketches: (i) mathematical nondimensionalization, (ii) state-variable choice, (iii) voltage scaling, and (iv) realization of the resulting coefficients by actual analog transfer functions. This separation is central to the verification strategy below.

\section{Lane--Emden benchmark family}

\subsection{Governing equation and global observables}

The Lane--Emden equation is
\begin{equation}
\frac{1}{\xi^2}\frac{d}{d\xi}\left(\xi^2\frac{d\theta}{d\xi}\right)+\theta^n=0,
\end{equation}
with regular center conditions $\theta(0)=1$ and $\theta'(0)=0$. Two global quantities are used throughout the project:
\begin{equation}
\theta(\xi_1)=0,
\end{equation}
and
\begin{equation}
\mu_1=-\xi_1^2\theta'(\xi_1).
\end{equation}
The first zero $\xi_1$ sets the dimensionless stellar surface and $\mu_1$ is the associated dimensionless mass coefficient.

\subsection{Conditioning of first-order state choices}

Several algebraically equivalent first-order forms were evaluated because analog hardware responds to signal magnitude and cancellation, not only mathematical equivalence. The raw flux-like state
\begin{equation}
u=-\xi^2\theta'
\end{equation}
behaves as $O(\xi^3)$ near the center and therefore produces a very small auxiliary voltage. The regularized state
\begin{equation}
v=\frac{u}{\xi^3}
\end{equation}
keeps the state $O(1)$ and replaces an $\xi^2$ denominator by $\xi$, but for $n=1$ it introduces a near-cancellation in $\theta-3v$. The derivative state
\begin{equation}
y=\theta'
\end{equation}
avoids both defects. The preferred physical convention is its positive counterpart
\begin{equation}
w=-\theta',
\end{equation}
which satisfies
\begin{align}
\frac{d\theta}{d\xi}&=-w,\\
\frac{dw}{d\xi}&=\theta^n-\frac{2w}{\xi}.
\end{align}

\begin{figure}[H]
\centering
\includegraphics[width=0.86\textwidth]{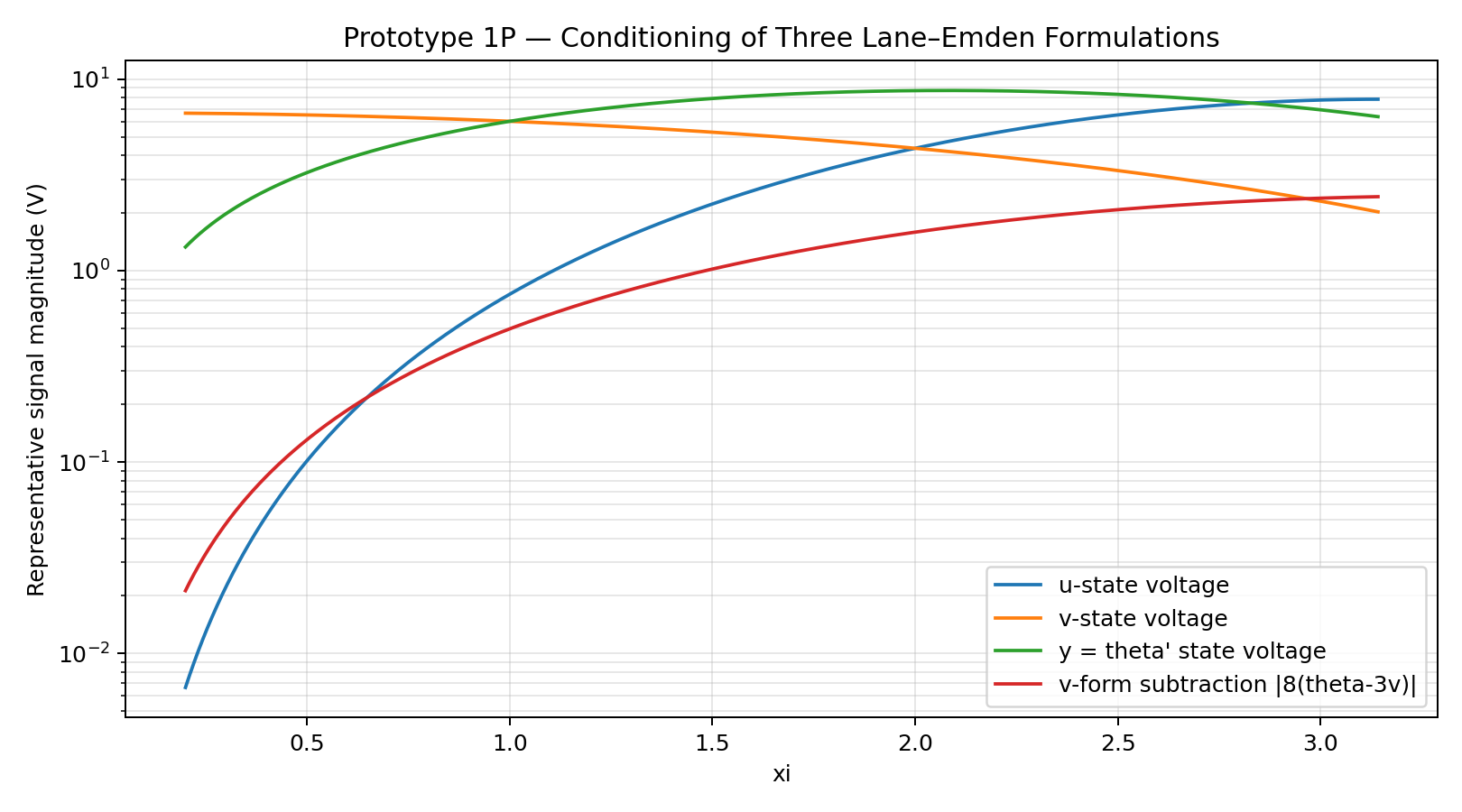}
\caption{Conditioning comparison among the raw $u$, regularized $v$, and derivative-state formulations. The preferred physical convention uses $w=-\theta'$ so that the auxiliary state remains positive.}
\label{fig:conditioning}
\end{figure}

At $\xi_0=0.2$, the original $u$-state implementation produced an auxiliary-state signal of only about $6.64$ mV under the then-used scaling, whereas the regularized $v$ state was about $6.64$ V. The derivative-state formulation avoids the small $u$ state and the $v$-form cancellation simultaneously. Figure~\ref{fig:conditioning} summarizes the comparison.

\subsection{Central regularity and finite-radius initialization}

The exact center cannot be connected directly to an electronic divider. The computation therefore starts at $\xi_0=0.2$ using a regular series. For general $n$,
\begin{align}
\theta(\xi)=&1-\frac{\xi^2}{6}+\frac{n\xi^4}{120}
+\frac{n(5-8n)\xi^6}{15120}\\
&+\frac{n(122n^2-183n+70)\xi^8}{3265920}+O(\xi^{10}),
\end{align}
with $w=-d\theta/d\xi$ obtained term-by-term. This converts the mathematical regularity condition into finite, directly settable state voltages.

Finite-radius initialization near a stellar center has an important historical analogue-computing precedent and is not claimed here as a new idea. In Kopal's 1949 Rockefeller differential-analyzer treatment of the standard $n=3$ polytropic model, reciprocal-radius factors also made direct operation near the origin inconvenient; local power-series solutions were used to set the analyzer's initial conditions at $x=0.1$ \citep{Kopal1949}. The distinction is computational: Kopal supplied tabulated $n=3$ Emden background functions to the machine and mechanically integrated the \emph{oscillation} equations, whereas the present benchmark circuit integrates the \emph{static Lane--Emden structure} itself and treats the start-state error as part of the calibrated electronic evidence chain.

\section{Reference solutions and ideal scaled-voltage verification}

The principal benchmark sequence uses $n=0$, $n=1$, and $n=3$. The first two have exact analytic solutions for the quantities used here; $n=3$ is referenced to a high-accuracy numerical integration. Table~\ref{tab:benchmarks} lists the adopted reference values and the later common-platform ideal results.

\begin{table}[H]
\centering
\caption{Principal polytropic benchmarks on the common ideal scaled-voltage platform.}
\label{tab:benchmarks}
\begin{tabular}{cccccc}
\toprule
$n$ & $\xi_1$ & $\mu_1$ & Run time (s) & Max. node (V) & $<9$ V \\
\midrule
0 & 2.449489743 & 4.898979486 & 7.498 & 8.981 & yes \\
1 & 3.141592654 & 3.141592654 & 9.805 & 7.947 & yes \\
3 & 6.896848619 & 2.018235951 & 22.323 & 7.947 & yes \\
\bottomrule
\end{tabular}
\end{table}

For $n=1$,
\begin{equation}
\theta=\frac{\sin\xi}{\xi},
\end{equation}
and $\xi_1=\mu_1=\pi$. For $n=0$, $\theta=1-\xi^2/6$, yielding $\xi_1=\sqrt{6}$ and $\mu_1=2\sqrt{6}$. The $n=3$ reference is $\xi_1=6.896848619\ldots$ and $\mu_1=2.018235951\ldots$, consistent with standard polytropic references \citep{Horedt2004}.

\begin{figure}[H]
\centering
\includegraphics[width=0.84\textwidth]{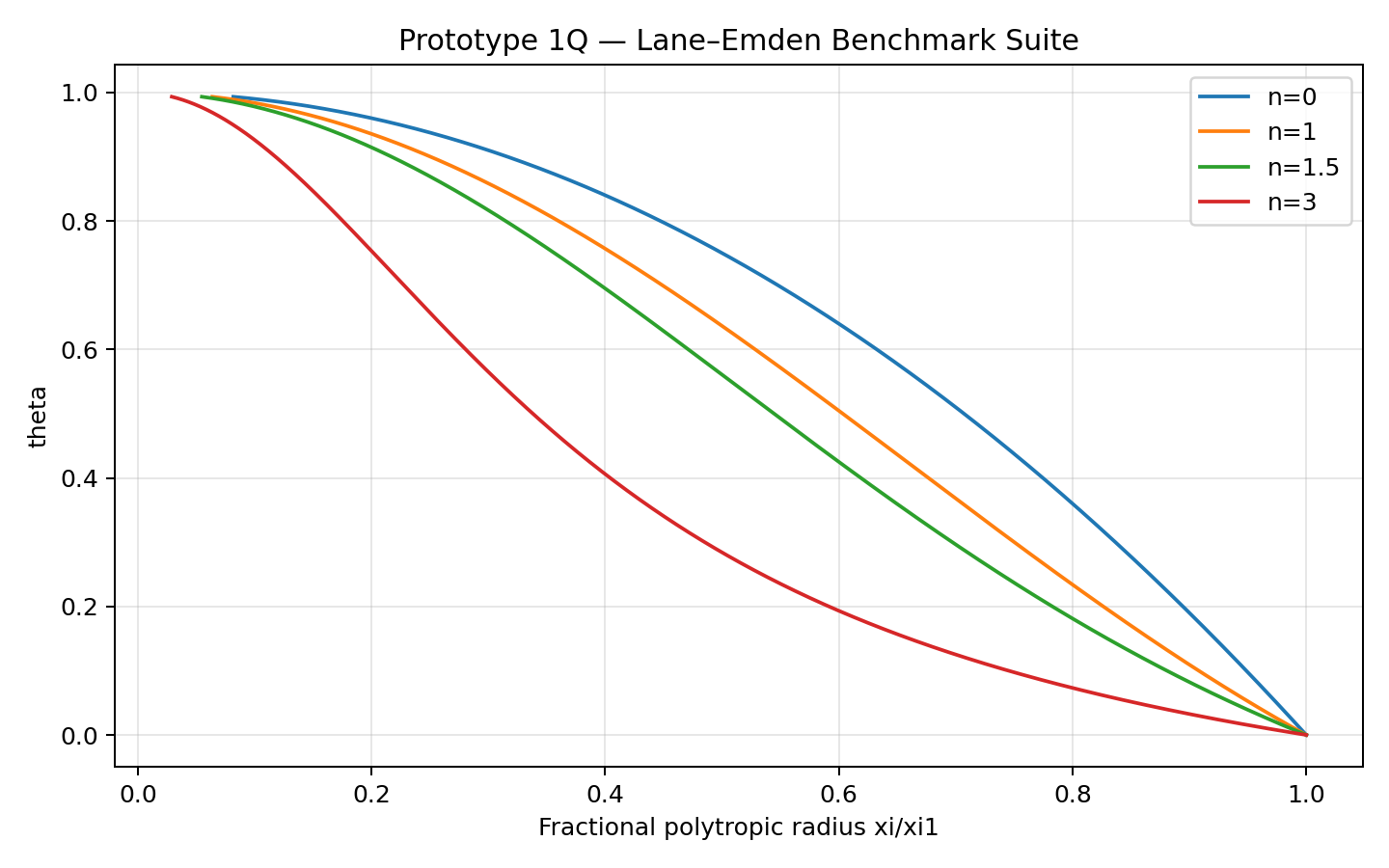}
\caption{Lane--Emden benchmark family used to define the analog stellar-structure verification problem. The $n=0$, $n=1$, and $n=3$ cases form the main hardware sequence; $n=1.5$ is retained as a possible later fractional-power test.}
\label{fig:profiles}
\end{figure}

Figure~\ref{fig:profiles} shows the benchmark-family profiles used in the numerical/reference stage.

\section{Hardware-oriented scaling and nonlinear source realization}

\subsection{Analog nonlinear blocks}

The hardware-oriented designs use explicit multiplier/divider transfer functions rather than abstract mathematical products. The primary nonlinear-block candidate is the AD734, whose transfer function is $W=XY/U$ and whose product documentation includes direct division and SPICE macromodels \citep{AD734}. Precision op-amp stages are based on the OPA197 family in the current vendor-model workflow; the device supports wide supply operation and precision low-bias-current integration compared with the 741-class exploratory circuits \citep{OPA197}.

These component choices are implementation candidates, not part of the Lane--Emden mathematics. The paper therefore reports equation coefficients separately from device realization and calibration.

\subsection{The $n=3$ headroom repair}

The $n=3$ benchmark requires a cubic source $\theta^3$. Two multiplier stages form the square and cube before a calibrated gain maps the raw multiplier output to the required source coefficient. An initial scaled design exceeded the conservative internal headroom target. Reducing the divider reference repaired one internal node but left the maximum slightly above 9 V. A second repair reduced the computation speed from the earlier value while preserving the target differential equation. The final case-specific design used $d\xi/d\tau=0.35$ and had a maximum ideal internal magnitude of 8.577 V, with numerical mapping errors of order $10^{-12}$.

This progression is intentionally retained as a methods result: computation speed provides an additional analog design degree of freedom when algebraic gain changes alone do not restore headroom.

\begin{figure}[H]
\centering
\includegraphics[width=0.86\textwidth]{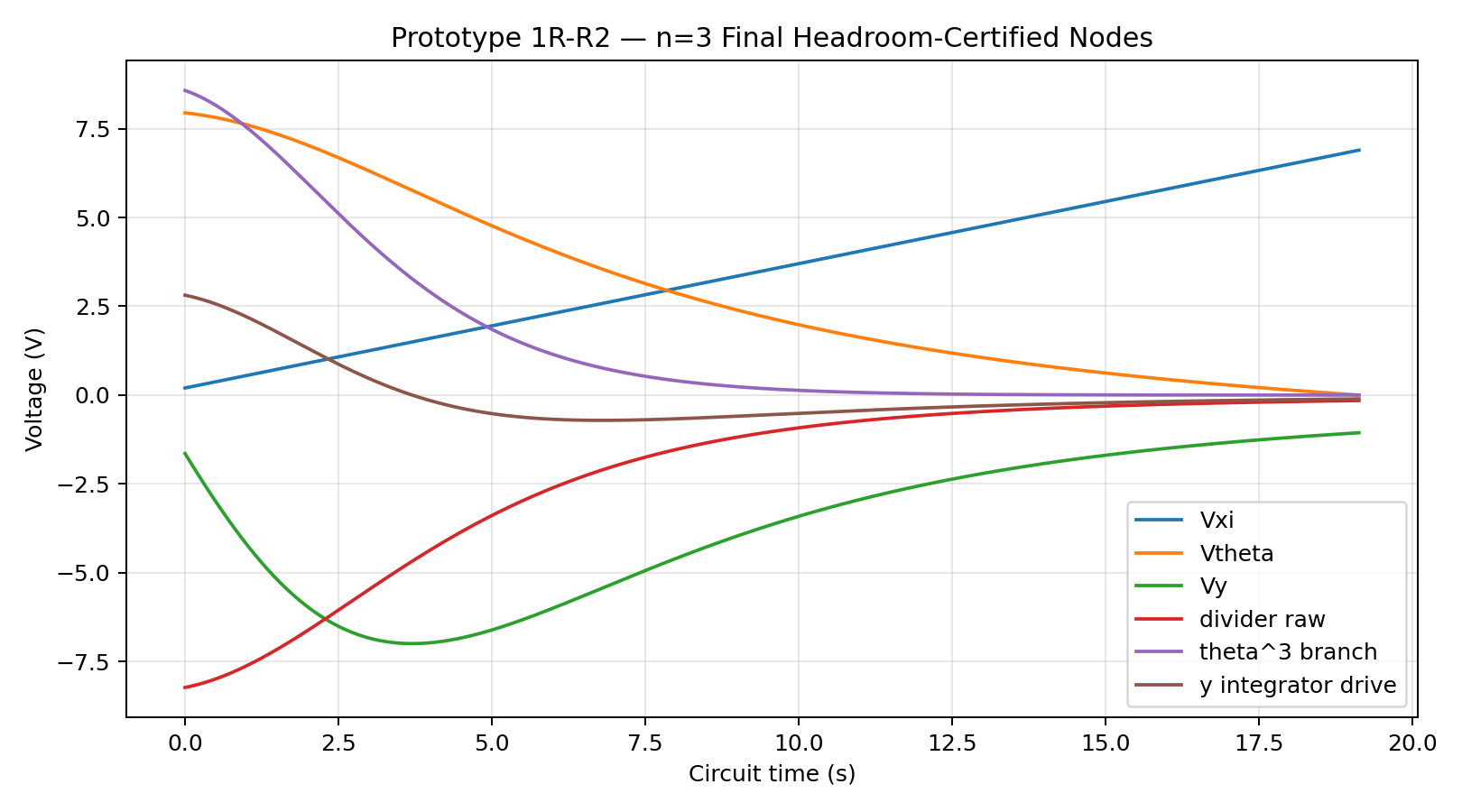}
\caption{Final case-specific $n=3$ ideal internal-node voltages after the coefficient-preserving headroom repairs.}
\label{fig:n3headroom}
\end{figure}

\subsection{A common $n=0,1,3$ hardware platform}

After the individual benchmark designs, a single common scaling was found:
\begin{equation}
V_\xi=\xi,\qquad V_\theta=8\theta,\qquad V_w=11w,
\end{equation}
with
\begin{equation}
\frac{d\xi}{d\tau}=0.30
\end{equation}
and a 1-V divider reference. The common theta-channel coefficient is $2.4/11$, and the divider branch uses a gain of 0.6. Only the source term changes:
\begin{align}
n=0:&\quad V_S=3.3~\mathrm{V},\\
n=1:&\quad V_S=0.4125V_\theta,\\
n=3:&\quad V_S=0.64453125W_3,
\end{align}
where $W_3$ is the raw two-multiplier cube output.

\begin{figure}[H]
\centering
\includegraphics[width=0.97\textwidth]{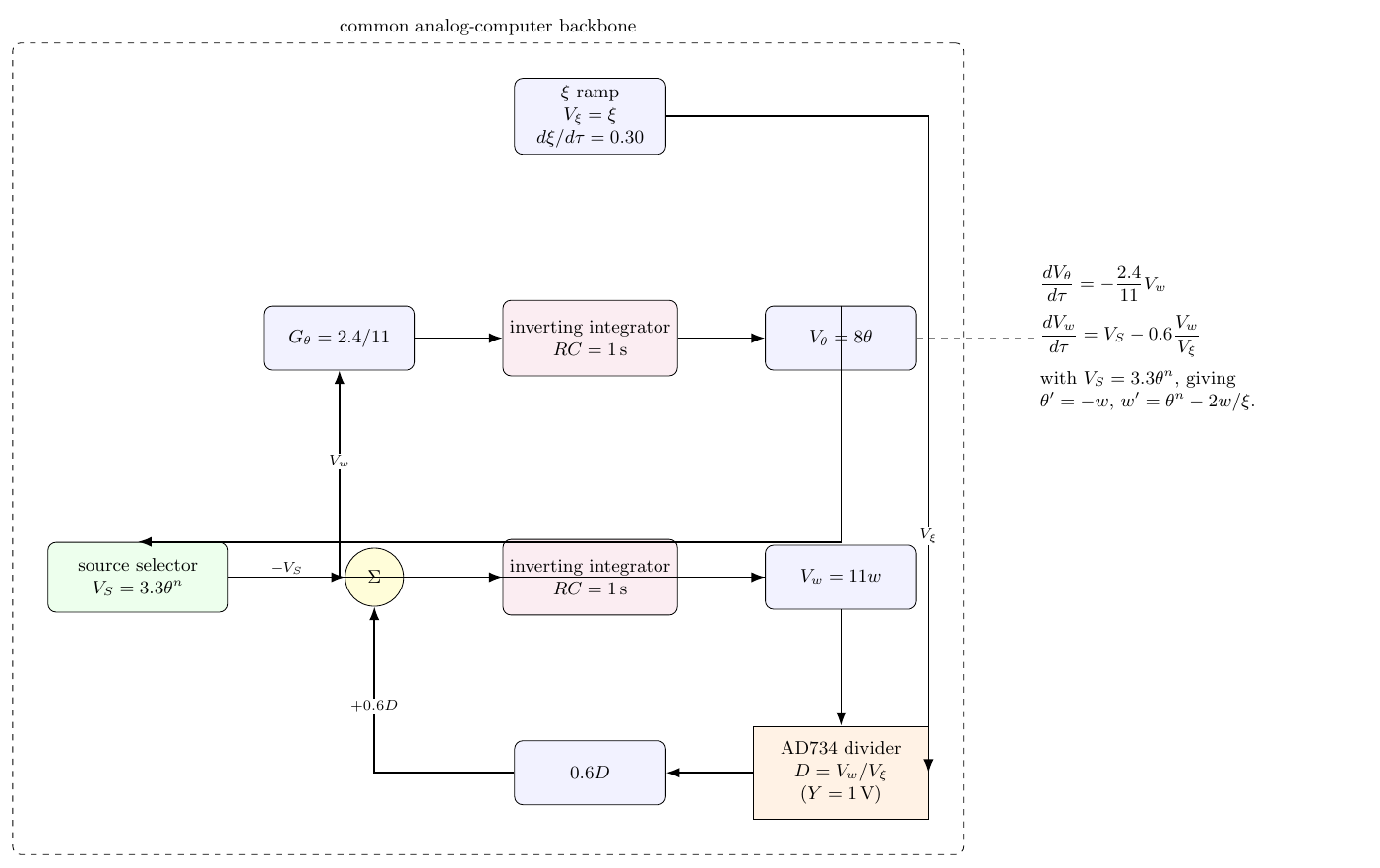}
\caption{Common positive-gradient ($w=-\theta'$) analog-computer architecture for the $n=0$, $n=1$, and $n=3$ Lane--Emden benchmarks. The inverting integrators implement the derivative sign explicitly; the AD734 divider realizes $V_w/V_\xi$ with a 1-V reference input.}
\label{fig:universalcircuit}
\end{figure}

This source-selector architecture permits the three principal benchmarks to share the same ramp, state integrators, and divider. The corresponding circuit-time equations are
\begin{align}
\frac{dV_\theta}{d\tau}&=-\frac{2.4}{11}V_w,\\
\frac{dV_w}{d\tau}&=V_S-0.6\frac{V_w}{V_\xi},
\end{align}
with $V_S=3.3\theta^n$. Figure~\ref{fig:universalcircuit} gives the common electronic signal flow. The nonlinear source is the only benchmark-dependent branch; its detailed selector implementation is given in Supplementary Section S2.1.

\begin{table}[H]
\centering
\caption{Common-platform voltage map and circuit-time equations. Coefficients are effective calibrated gains; they are reported independently of any particular resistor-network realization.}
\label{tab:commonmapping}
\begin{tabular}{p{0.185\textwidth}p{0.22\textwidth}p{0.27\textwidth}p{0.22\textwidth}}
\toprule
Quantity & Voltage mapping & Required circuit-time relation & Principal analog block \\ 
\midrule
Independent variable & $V_\xi=\xi$ & $dV_\xi/d\tau=0.30$ V s$^{-1}$ & calibrated ramp \\ 
Structure state & $V_\theta=8\theta$ & $dV_\theta/d\tau=-(2.4/11)V_w$ & gain + inverting integrator \\ 
Gradient state & $V_w=11w$ & $dV_w/d\tau=V_S-0.6(V_w/V_\xi)$ & source, divider, summer, integrator \\ 
Polytropic source & $V_S=3.3\theta^n$ & $n=0,1,3$ selectable & constant / gain / multiplier chain \\ 
\bottomrule
\end{tabular}
\end{table}

\begin{figure}[H]
\centering
\includegraphics[width=0.84\textwidth]{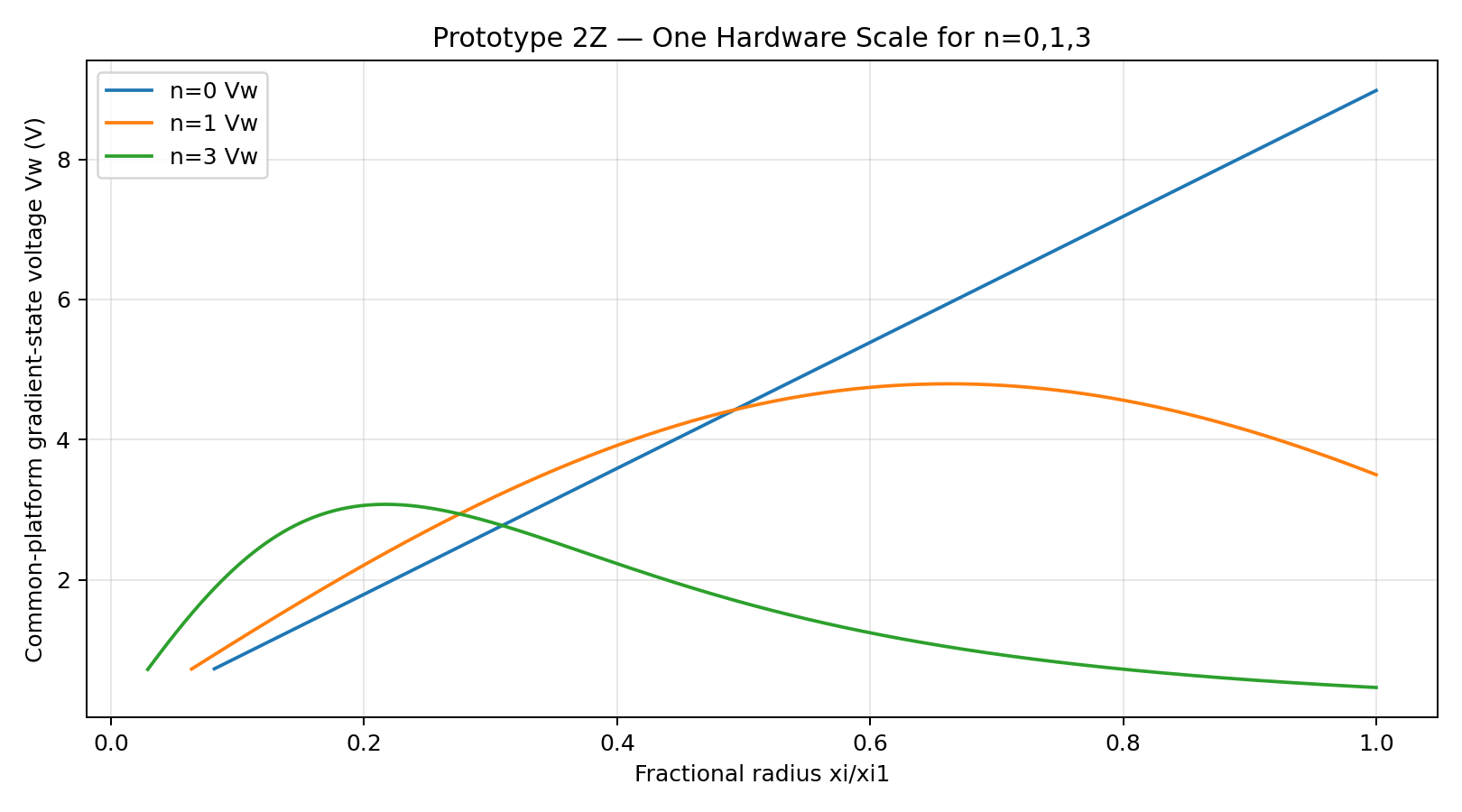}
\caption{Common-platform gradient-state voltage for the $n=0$, $n=1$, and $n=3$ benchmark family.}
\label{fig:commonplatform}
\end{figure}

Figure~\ref{fig:commonplatform} shows the resulting auxiliary-state voltages.

\section{Verification hierarchy}

The project deliberately distinguishes evidence levels rather than treating all simulations as equivalent:
\begin{enumerate}
\item \textbf{Mathematical reference:} exact or high-accuracy Lane--Emden solutions and independently integrated stellar test systems.
\item \textbf{Ideal scaled-voltage model:} direct integration of the equations produced by the proposed voltage mapping.
\item \textbf{Functional SPICE:} circuit-topology netlists using behavioral functional blocks.
\item \textbf{Vendor-macromodel SPICE:} explicit device models, finite supplies, initialization, and component tolerances.
\item \textbf{Physical hardware:} calibrated measured waveforms, repeatability, and uncertainty propagation.
\end{enumerate}

Mathematical and ideal scaled-voltage verification are complete for the main polytropic benchmarks. Functional-SPICE decks have been generated for the preferred common architecture, but vendor-level execution of that preferred $w$-form platform remains a separate gate. The regularized $v$-form device-level realization has now completed calibrated-active vendor-macromodel simulation certification under the predefined engineering criterion; the result is summarized below and has been independently recomputed from the archived raw CSV tables. Physical hardware remains pending.

This distinction is important because an accurate ideal ODE integration does not, by itself, establish a realizable analog electronic computer.

\section{Sensitivity, Monte Carlo propagation, and calibration}

\subsection{Local effective-parameter sensitivities}

For a stellar observable $Q$ and effective calibration parameter $p$, the local sensitivity is defined as
\begin{equation}
S_{Q,p}=\frac{\partial\ln Q}{\partial\ln p}.
\end{equation}
The uncertainty model operates on installed effective quantities such as ramp scale, integrator time constants, gain ratios, divider scale, source-block gain, nonlinear-power gain, and initial conditions. It therefore does not equate a raw resistor tolerance directly with a stellar-output uncertainty.

A central result is that the nonlinear $n=3$ benchmark is more sensitive than $n=0$ or $n=1$ to several effective errors. For example, the local $n=3$ sensitivity of $\xi_1$ to the divider-branch gain is approximately 1.585, while the $\mu_1$ sensitivity to ramp speed is approximately 3.022.

\subsection{Monte Carlo propagation}

Independent bounded effective errors were propagated through 2000 realizations per benchmark. Table~\ref{tab:mc} reports the standard deviations in the two global stellar quantities.

\begin{table}[H]
\centering
\caption{Monte Carlo output dispersion for independent bounded effective calibration errors. These are design-stage uncertainty studies, not measured hardware distributions.}
\label{tab:mc}
\begin{tabular}{ccccc}
\toprule
$n$ & Bounded error & $\sigma(\xi_1)$ & $\sigma(\mu_1)$ & Max. $|\Delta\mu_1|$ \\
 & (\%) & (\%) & (\%) & (\%) \\
\midrule
0 & 0.1 & 0.071 & 0.178 & 0.498 \\
1 & 0.1 & 0.062 & 0.179 & 0.523 \\
3 & 0.1 & 0.144 & 0.243 & 0.806 \\
0 & 1.0 & 0.716 & 1.790 & 5.750 \\
1 & 1.0 & 0.662 & 1.830 & 5.376 \\
3 & 1.0 & 1.437 & 2.388 & 6.993 \\
\bottomrule
\end{tabular}
\end{table}

\begin{figure}[H]
\centering
\includegraphics[width=0.84\textwidth]{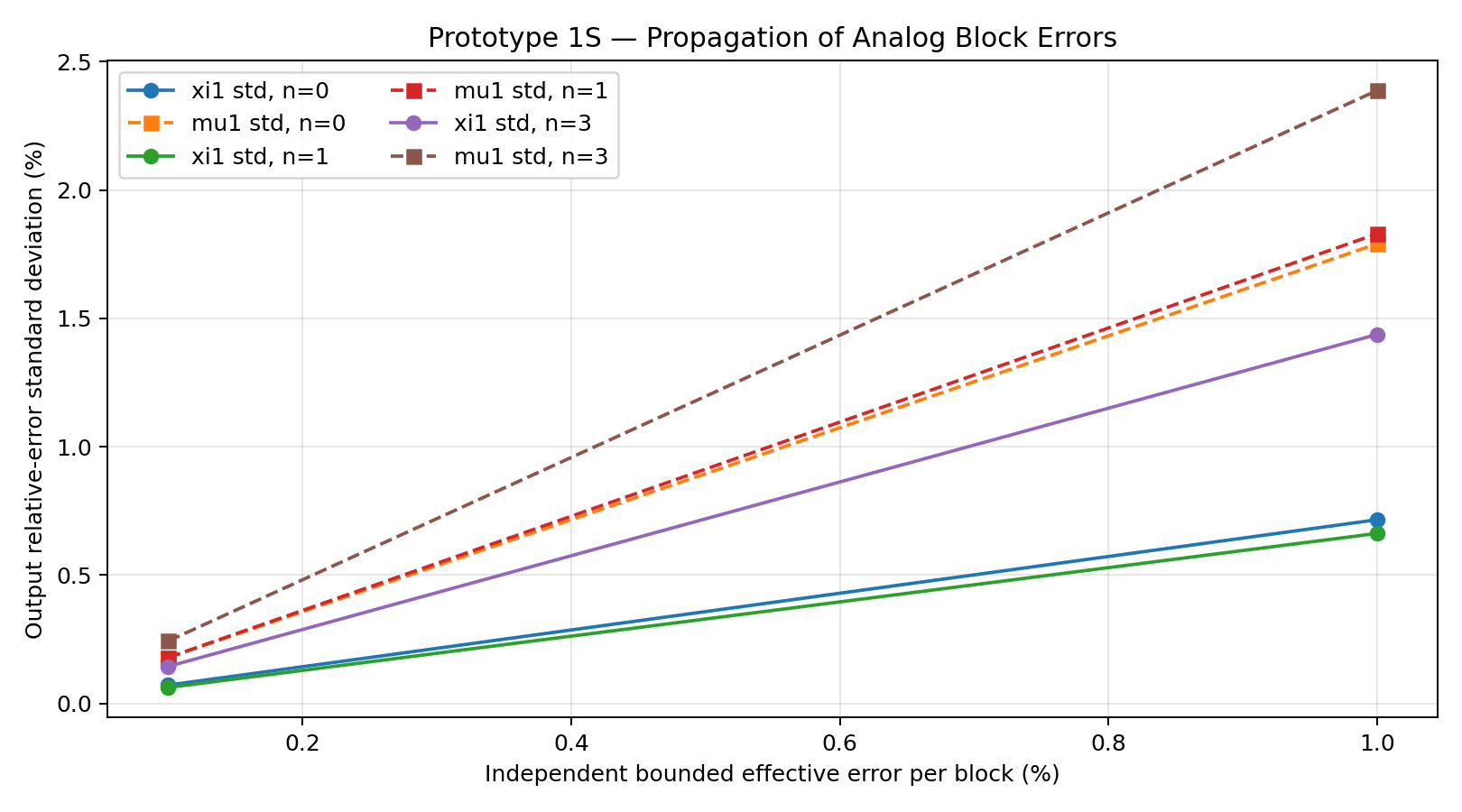}
\caption{Propagation of effective block errors to the global polytropic observables.}
\label{fig:errorprop}
\end{figure}

\subsection{Calibration allocation}

The design-stage calibration allocation targets post-calibration effective uncertainties, not raw component specifications. The selected allocation was constructed to keep the first-order RMS uncertainty in both $\xi_1$ and $\mu_1$ below approximately 0.1\% for $n=0,1,3$. The resulting predicted dispersions are
\begin{center}
\begin{tabular}{ccc}
\toprule
$n$ & $\sigma(\xi_1)$ (\%) & $\sigma(\mu_1)$ (\%) \\
\midrule
0 & 0.0364 & 0.0767 \\
1 & 0.0344 & 0.0755 \\
3 & 0.0691 & 0.0990 \\
\bottomrule
\end{tabular}
\end{center}

\begin{figure}[H]
\centering
\includegraphics[width=0.82\textwidth]{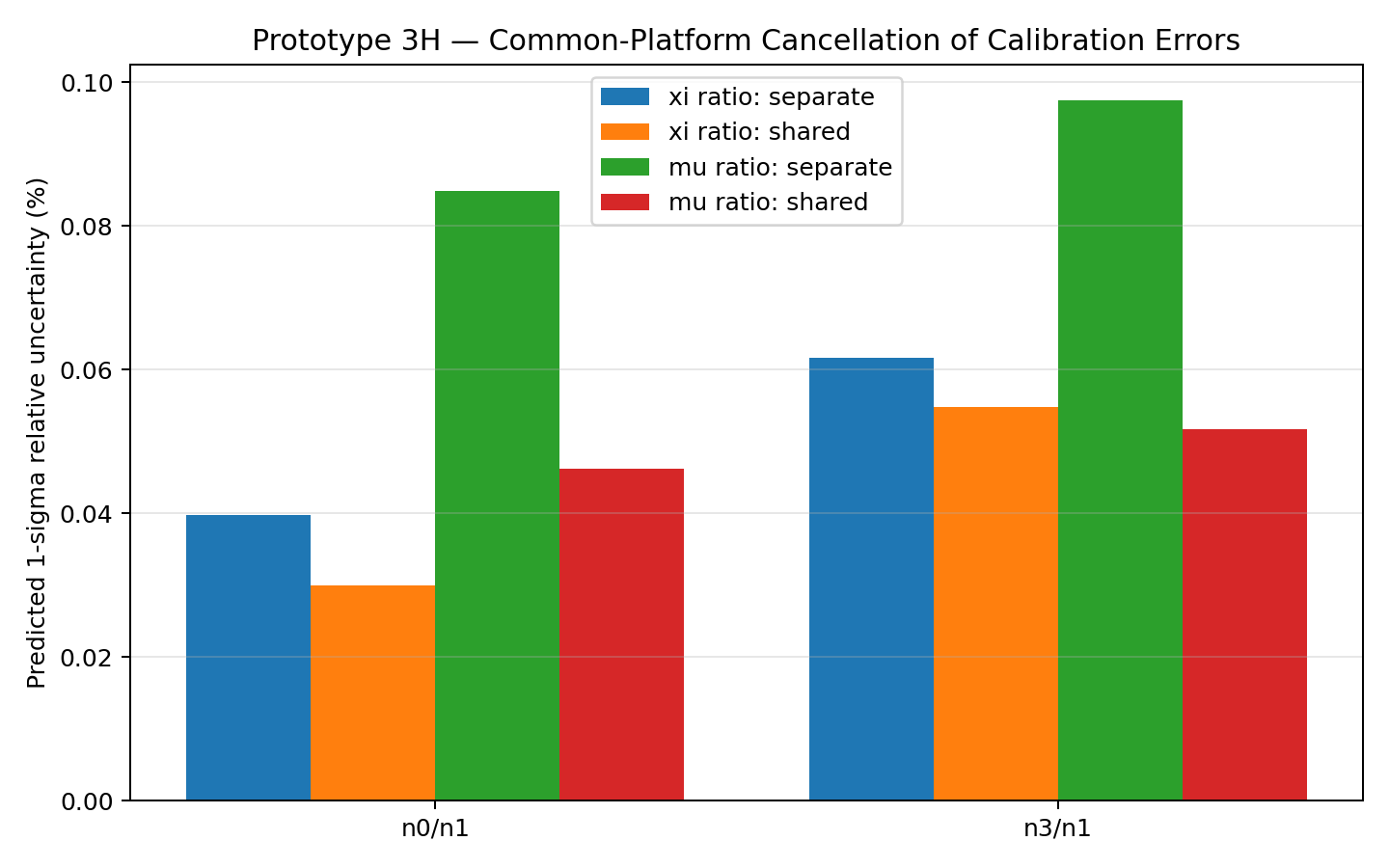}
\caption{Predicted reduction in cross-benchmark ratio uncertainty when the principal polytropes share one calibrated analog-computer backbone. Absolute uncertainties remain primary and are not replaced by the ratio measurement.}
\label{fig:commonmode}
\end{figure}

The common-platform architecture also introduces useful error correlation. In the design study, the predicted uncertainty of the ratio $\mu_1(n=3)/\mu_1(n=1)$ decreased from about 0.0976\% for separately calibrated boards to about 0.0517\% when shared calibration parameters were treated as common. Figure~\ref{fig:commonmode} illustrates this differential-metrology advantage.

\section{Dimensionless self-closed stellar-structure extension}

Polytropes isolate the numerical/electronic problem, but they do not exercise a closed set of mass, pressure, luminosity, and temperature equations simultaneously. A separate dimensionless architecture was therefore constructed as a four-state demonstration:
\begin{align}
q' &= 3x^2 d,\\
p' &= -\frac{qd}{x^2},\\
\ell' &= x^2d\hat\epsilon,\\
\theta' &= -\frac{\ell\hat\kappa d}{x^2\theta^3},
\end{align}
with closures
\begin{equation}
d=\frac{p}{\theta},\qquad
\hat\kappa=d\theta^{-7/2},\qquad
\hat\epsilon=d\theta^4.
\end{equation}
At $x=1$, the reference integration gives approximately
\begin{equation}
(q,p,\ell,\theta)=(0.8187733,0.6135453,0.1627310,0.8642618).
\end{equation}
The numerical implementation converges systematically with step refinement; at 2001 grid points the maximum endpoint discrepancy against the high-accuracy reference is below $10^{-8}$ and decreases further with refinement.

\begin{figure}[H]
\centering
\includegraphics[width=0.84\textwidth]{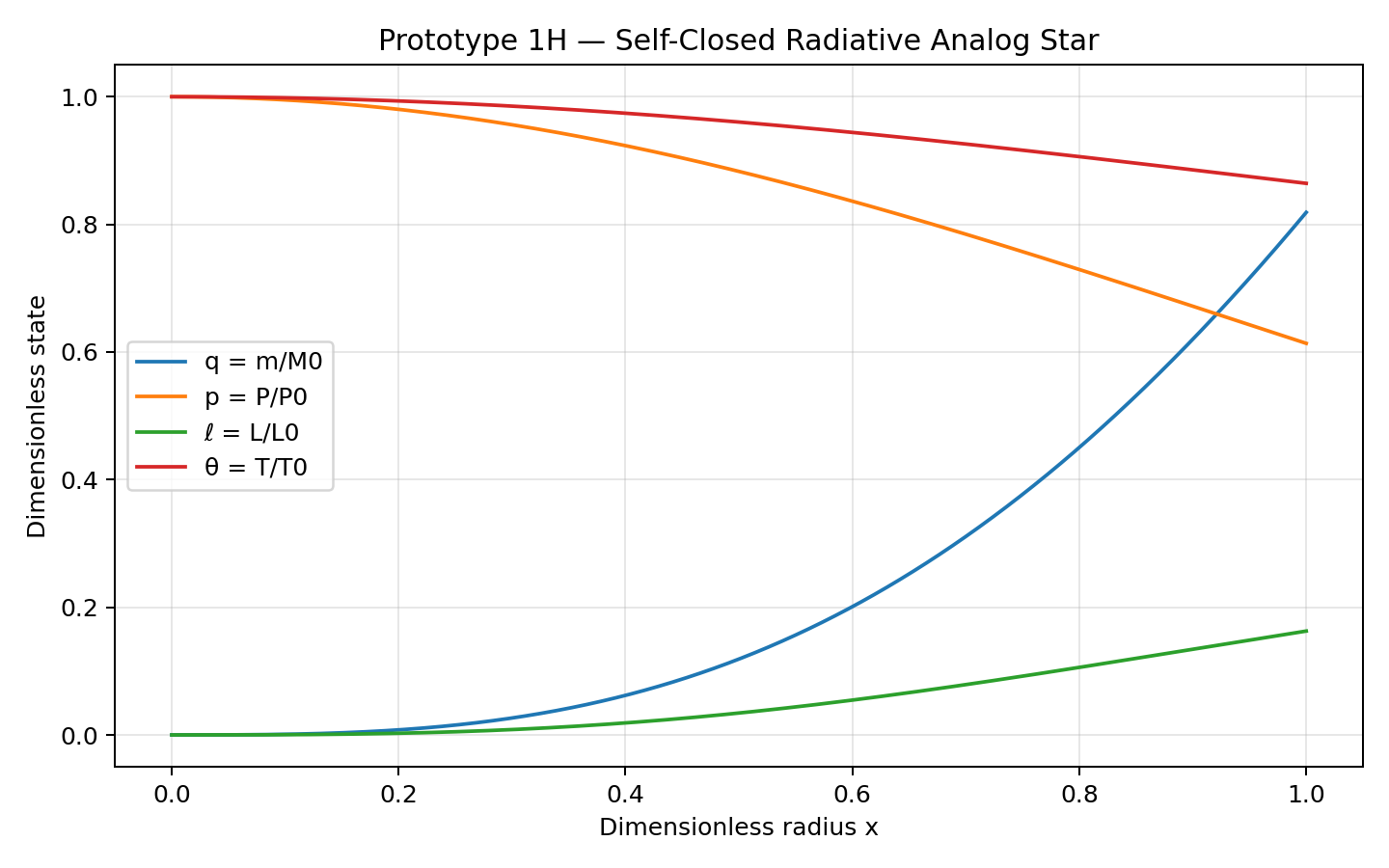}
\caption{Dimensionless self-closed four-state stellar-structure architecture. This system is an architecture demonstration with idealized closures and is not interpreted as a calibrated physical star.}
\label{fig:selfclosed}
\end{figure}

A hardware realization of the opacity law $\theta^{-7/2}$ requires a genuine function generator, log/antilog stage, calibrated approximation, or equivalent nonlinear block. It cannot be replaced by a chain of ordinary multipliers without changing the modeled function.

\section{Vendor-macromodel certification result}
\label{sec:1orcert}

The device-level certification workflow was closed first on the regularized
$v$-form infrastructure because that branch already contained explicit
AD734/OPAx197 integration. The governing equations, voltage scaling, and nominal time constant were held fixed during tolerance-model repair:
\begin{equation}
\theta'=-\xi v,\qquad
v'=\frac{\theta-3v}{\xi},
\end{equation}
with $V_\xi=2.5\xi$, $V_\theta=8\theta$, $V_v=20v$,
$d\xi/d\tau=0.5$, $\xi_0=0.2$, and nominal $RC=1$ s.

The original uncalibrated tolerance envelope contained deterministic root
failures. The calibrated regularized-$v$ realization therefore modifies only the tolerance implementation:
the two integrator time constants are matched and trimmed, multiplier/divider
branches are represented by installed affine calibration residuals, and the
two integrator bias terms are independently nulled. The predefined
engineering certification criterion was kept fixed at
\begin{equation}
|\xi_1-\pi|\le 0.020,
\label{eq:1orgate}
\end{equation}
with no saturation or convergence failure permitted. The tighter 0.1\%
relative first-zero target is reported separately and does not define the
certification pass/fail decision.

\begin{figure}[H]
\centering
\includegraphics[width=0.99\textwidth]{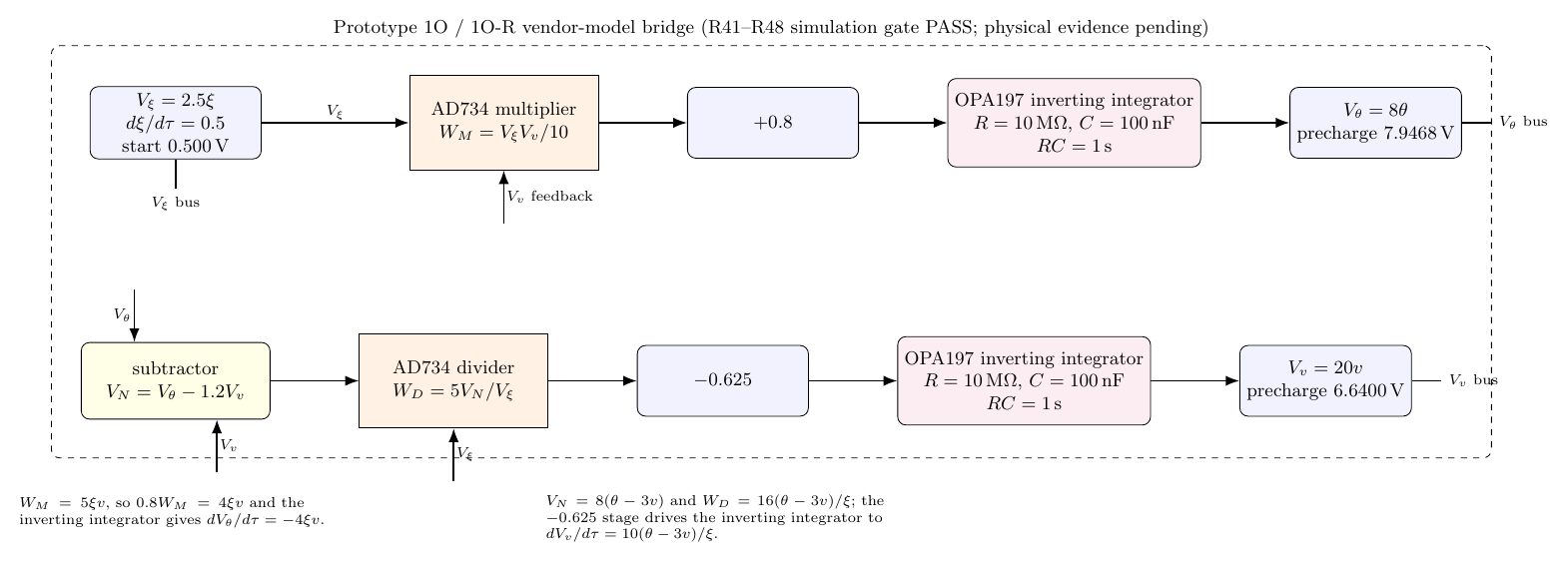}
\caption{Regularized-$v$ $n=1$ vendor-macromodel bridge. The fixed
$n=1$ equations are $\theta'=-\xi v$ and
$v'=(\theta-3v)/\xi$. The calibrated-active simulation retains
$R=10$ M$\Omega$, $C=100$ nF, and nominal $RC=1$ s while assigning
calibrated residuals only to the authorized tolerance layer.}
\label{fig:vform1O}
\end{figure}

\subsection{Deterministic certification}

The final calibrated-active deterministic suite contains 67 declared
certification cases. Independent reduction of the returned case table
reproduces the stored gate flags directly from $\xi_1$: all 67 cases converge,
all 67 satisfy Eq.~\eqref{eq:1orgate}, and no OPA saturation or AD734
full-scale event occurs. The worst certified error is
\begin{equation}
\max|\xi_1-\pi|=0.019745733,
\end{equation}
leaving only $2.54267\times10^{-4}$ of margin to the engineering threshold.
Thus the pass is real but not loose.

\begin{table}[H]
\centering
\caption{Independent recomputation of the regularized-$v$ deterministic
vendor-macromodel certification. The 0.1\% column is characterization only.}
\label{tab:1ordeterministic}
\begin{tabular}{lr}
\toprule
Quantity & Independent result \\
\midrule
Declared certification cases & 67 \\
Engineering-gate pass & 67/67 \\
Convergence & 67/67 \\
OPA saturation events & 0 \\
AD734 full-scale exceedances & 0 \\
Worst $|\xi_1-\pi|$ & 0.019745733 \\
Remaining gate margin & 0.000254267 \\
Strict 0.1\% relative $\xi_1$ & 11/67 \\
Deterministic $\mu_1$ range & 3.139093060--3.222052035 \\
\bottomrule
\end{tabular}
\end{table}

Ten uncalibrated negative controls are preserved outside the repaired
certification population. All ten converge and remain unsaturated; four
reproduce the historical root-gate failures. Their persistence is useful:
the successful certification result arises from the declared calibration layer rather than from
weakening the raw datasheet-stress controls.

\begin{figure}[H]
\centering
\includegraphics[width=0.88\textwidth]{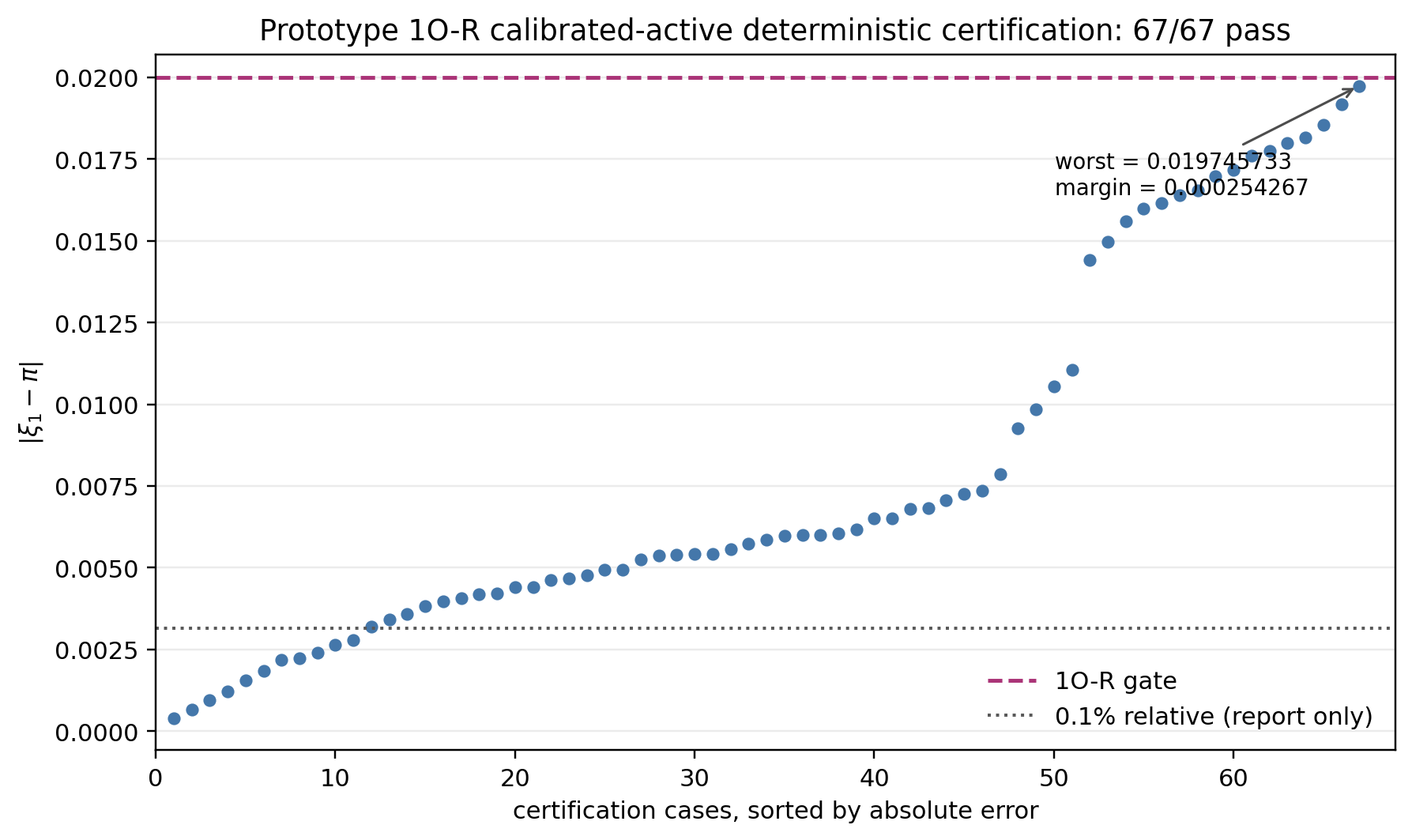}
\caption{The 67 calibrated-active deterministic certification cases, sorted
by absolute first-zero error. The upper dashed line is the predefined
engineering certification threshold. The much tighter 0.1\% relative level is report-only.}
\label{fig:1ordeterministic}
\end{figure}

\subsection{Seeded Monte Carlo certification}

The final certification population uses passive seed 1010734197 and an independent
active-residual seed 1041048. Manufacturer datasheet worst-case bounds are
not converted into process distributions; active devices remain at the
official nominal macromodel in the Monte Carlo population, while the declared
passive, trim, thermal-tracking, gain-residual, and bias-residual engineering
allocations are sampled.

\begin{table}[H]
\centering
\caption{Independent reduction of the final regularized-$v$
Monte Carlo populations. Each phase contains 1000 cases.}
\label{tab:1ormc}
\begin{tabular}{lcccc}
\toprule
Phase & Gate pass & Max. $|\xi_1-\pi|$ & Strict 0.1\% & $\mu_1$ range \\
\midrule
Pre-trim $\pm0.50\%$ & 1000/1000 & 0.017604792 & 274/1000 & 3.142388--3.217420 \\
Preferred post-trim $\pm0.20\%$ & 1000/1000 & 0.013136114 & 121/1000 & 3.154406--3.201666 \\
Stretch post-trim $\pm0.10\%$ & 1000/1000 & 0.012611280 & 90/1000 & 3.153984--3.203186 \\
\bottomrule
\end{tabular}
\end{table}

All 3000 Monte Carlo cases converge and none saturates. For 1000 successes and
zero failures in a phase, the exact one-sided 95\% lower binomial bound on the
pass probability is 0.99700875 (99.7009\%) for the declared engineering
population. This statement does not imply knowledge of an undisclosed
manufacturer process distribution.

The lower fraction meeting the report-only 0.1\% target after trimming is not
a contradiction. The nominal vendor model has a positive first-zero bias, so
some of the broader negative pre-trim mismatch draws accidentally cancel that
bias. Centering the installed $RC$ ratio removes that chance cancellation
while improving the intended calibrated realization.

\begin{figure}[H]
\centering
\includegraphics[width=0.82\textwidth]{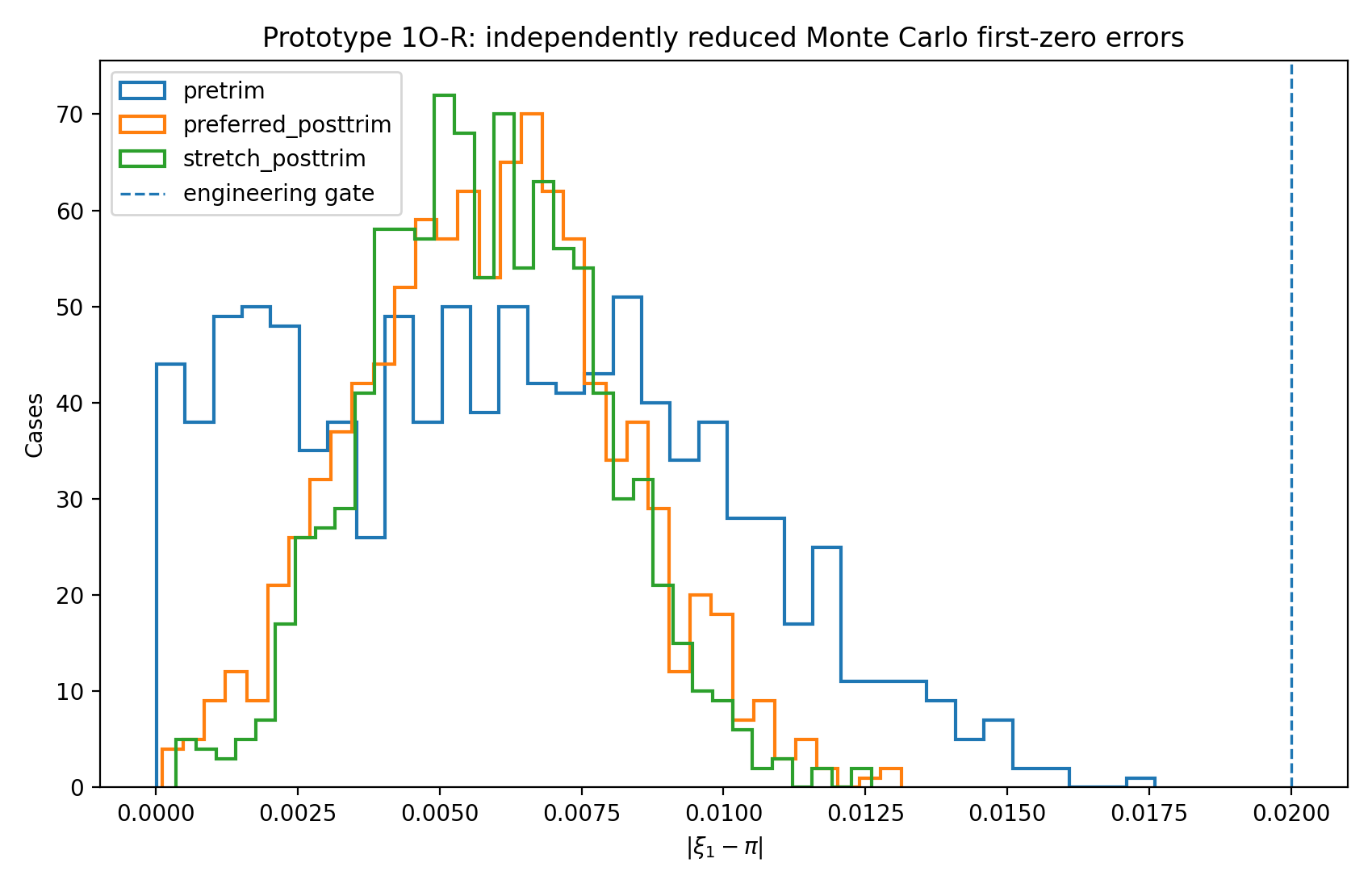}
\caption{Independent reduction of the three final 1000-case
Monte Carlo populations. Every case lies below the 0.020 engineering gate.
The figure was regenerated directly from the archived serialized CSV tables rather than from
the original plotting workflow.}
\label{fig:1ormcaudit}
\end{figure}

\subsection{Observable characterization and evidence recovery}

The preferred post-trim population also reproduces the preregistered
diagnostic ordering: differential $RC$ ratio is the leading linear predictor
of $\xi_1$ ($r^2=0.284902$), while common mean $RC$ is the leading predictor
of $\mu_1$ ($r^2=0.782800$). These Pearson rankings are characterization, not
a Sobol or variance-decomposition claim.

The completed Monte Carlo process experienced a serialization interruption:
a preserved partial file contained 2278 rows (1000 preferred, 1000 pre-trim,
278 stretch). The missing stretch tail was regenerated with the same seeds.
Independent comparison shows that every preserved overlap row maps one-to-one
onto the final 3000-row file with zero serialized difference in $\xi_1$,
$\mu_1$, OPA headroom, and AD734 headroom. The final file has 3000 unique
phase/pair-index rows and its SHA-256 matches the recovery record. In
addition, all 2336 entries in the returned package manifest pass independent
SHA-256 verification.

\begin{figure}[H]
\centering
\includegraphics[width=0.86\textwidth]{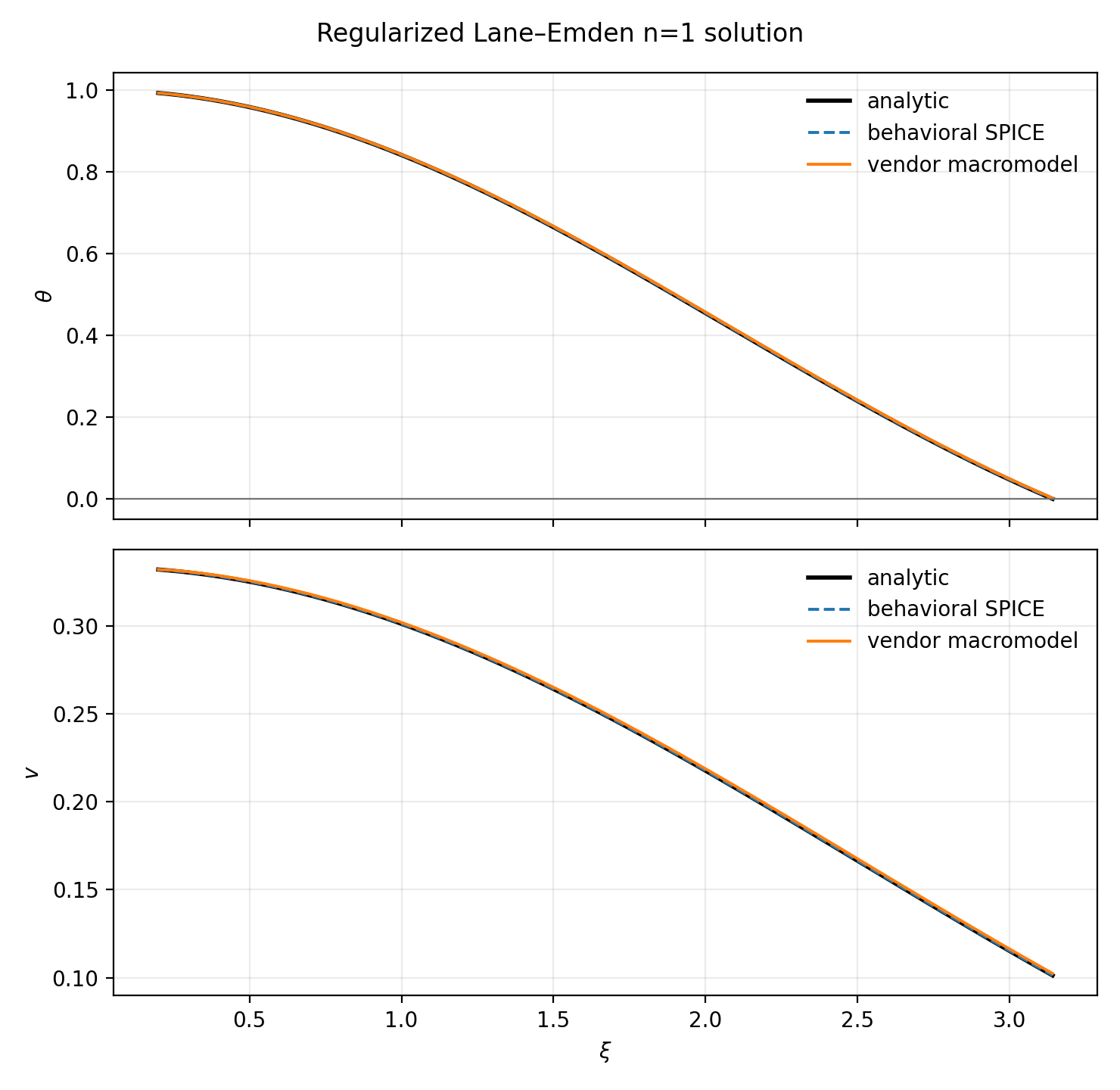}
\caption{Nominal $n=1$ regularized-$v$ solution used as the device-level
bridge: analytic reference, behavioral SPICE, and official vendor-macromodel
trajectory. The behavioral result is a scaling reference; vendor certification
is provided by the calibrated-active deterministic and Monte Carlo evidence,
not by visual agreement of this nominal waveform alone.}
\label{fig:1ornominal}
\end{figure}

\subsection{Certification boundary}

The independently audited result closes the regularized-$v$
\emph{vendor-macromodel simulation} robustness gate for the declared
calibrated-active tolerance allocation. It does not certify a physical trim
network or authorize a hardware-accuracy claim. Installed range, resolution,
temperature drift, branch offsets, ramp accuracy, leakage, loading, noise,
aging, parasitics, and bench repeatability remain physical-evidence
obligations. Likewise, an alternative residual-allocation study considered during development
was not used as a requirement for the present certification.

\section{Discussion}

\subsection{Mathematical variable choice is an electronics-design choice}

The progression from $u$ to $v$ to derivative-state formulations demonstrates that algebraically equivalent equations can have very different analog realizability. The central scaling of an auxiliary variable determines measurable voltage amplitude; subtraction structure determines cancellation; and the independent-variable power in a divider determines sensitivity near the center. For analog scientific computing, variable choice is therefore part of hardware conditioning rather than a purely symbolic preference.

\subsection{Historical differential-analyzer precedent clarifies the contribution}

Rosseland's Oslo Analyzer and Kopal's 1949 calculation make the historical boundary especially clear. The Oslo machine placed large-scale mechanical differential-equation computation inside an institute devoted to theoretical astrophysics \citep{Holst1996}, and Kopal described Rosseland's work as a pioneer application of the differential analyzer to astronomical problems \citep{Kopal1949}. Kopal's own $n=3$ stellar-oscillation setup used nine differential-analyzer integrators, supplied precomputed Emden background functions through input tables, transformed products into integrable machine form, and used a local series to avoid excessive reciprocal-radius scaling near the origin. The present work therefore does not claim originality for analogue computation in astrophysics, analogue stellar computation, center-series initialization, or machine-oriented algebraic reformulation. Its narrower contribution is the verified component-level \emph{electronic} realization of the static Lane--Emden benchmark family, together with common voltage scaling, nonlinear source selection, global-observable checks, vendor-macromodel certification of a device-level bridge, and an explicit calibration/uncertainty hierarchy.

\subsection{Computation speed provides headroom without changing the stellar solution}

The $n=3$ repair shows that the time-to-$\xi$ mapping can be used to relieve voltage stress while preserving the intended dimensionless trajectory. This is analogous to rescaling the independent variable in a digital ODE solver, but in hardware it directly changes gain requirements and saturation margin.

\subsection{Calibration matters more than nominal tolerance}

The error studies consistently favor calibration of installed effective blocks. A nominally precise resistor does not by itself specify the effective integrator time constant, and a multiplier headline accuracy does not determine the fitted scale, offset, and trajectory-dependent residual of the installed nonlinear branch. The manuscript therefore treats component tolerances as inputs to calibration design rather than as substitutes for calibration.

\subsection{Derived constraints for an eventual physical implementation}

Although physical hardware is not yet part of the evidence base, the certified mathematical and sensitivity models already impose quantitative design constraints on a future bench implementation. Table~\ref{tab:futureconstraints} collects the principal values presently used for hardware planning. These are design/calibration targets rather than measured hardware specifications. They define future physical obligations and do not retroactively change the closed vendor-macromodel simulation engineering gate.

\begin{table}[H]
\centering
\caption{Derived design constraints for a future physical implementation. Values are project-level engineering/calibration targets, not measured hardware performance.}
\label{tab:futureconstraints}
\begin{tabular}{p{0.31\textwidth}p{0.22\textwidth}p{0.35\textwidth}}
\toprule
Quantity & Current target/screen & Interpretation \\
\midrule
Differential integrator-$RC$ residual & $\lesssim 0.20\%$ preferred & Post-trim target for first-zero accuracy margin \\
Mean/common $RC$ calibration & $\lesssim 0.05\%$ & Controls absolute timing scale, especially $\mu_1$ \\
$V_\xi$ ramp-scale calibration & $\lesssim 0.02\%$ & Separate from common $RC$ because the waveform alone cannot identify them independently \\
$V_\theta$ precharge & $\sim 7.8$ mV single-parameter 0.1\% screen & Most sensitive initialization voltage in the present $v$-form analysis \\
$V_\xi$ start value & $\sim 37$ mV single-parameter 0.1\% screen & Corresponds to roughly 30 ms at the nominal 1.25 V s$^{-1}$ ramp \\
Effective parallel leakage, $\theta$ state & $\gtrsim 49$ G$\Omega$ 0.1\% surrogate screen & Combined switch, board, capacitor, and amplifier leakage must be assessed as installed drift \\
Effective parallel leakage, $v$ state & $\gtrsim 9.5$ G$\Omega$ 0.1\% surrogate screen & Less restrictive than the $\theta$ state in the present sensitivity model \\
Waveform acquisition & 16-bit, 50--100 Hz, $\sim0.5$ mV RMS representative study & Gives sub-0.01\% RMS first-zero reduction error in the current synthetic acquisition screen \\
Trigger synchronization & 1 ms design target & Leaves large margin relative to the present $V_\xi$ start-time tolerance \\
\bottomrule
\end{tabular}
\end{table}

These values also explain why the eventual board layout must be treated as part of the numerical method. Ten-megaohm integrator nodes are sensitive to switch leakage, board contamination, humidity, probing, and surface conduction. Conversely, dynamic slew is not expected to be limiting: the ideal $n=1$ waveforms require only order-unity volts per second, many orders of magnitude below the slew capability of modern precision operational amplifiers. The future hardware program therefore prioritizes leakage control, calibration stability, and reproducible acquisition over bandwidth.

\subsection{Reproducibility architecture and traceability}

The project treats reproducibility as part of the numerical method rather than as a post-processing step. Each major result is associated with an equation definition, a scaling definition, a machine-readable output, a reduction script, and an evidence label. Table~\ref{tab:traceability} summarizes the minimum traceability chain that will be required in the submission archive. This structure is intended to make it possible to distinguish a change in the stellar equations from a change in circuit scaling, component realization, calibration, or waveform reduction.

\begin{table}[H]
\centering
\caption{Minimum traceability chain for the submission archive.}
\label{tab:traceability}
\begin{tabular}{p{0.25\textwidth}p{0.29\textwidth}p{0.35\textwidth}}
\toprule
Artifact class & Required content & Purpose \\
\midrule
Equation specification & state definitions, signs, coefficients, initial conditions & fixes the mathematical problem independently of electronics \\
Scaling specification & voltage scales, computation speed, nominal gains, reference conventions & fixes the equation-to-voltage map \\
Circuit/netlist & ideal or vendor-level topology with version/hash & fixes the electronic realization \\
Calibration record & installed $RC$, ramp, nonlinear-branch scale/offset, temperature & separates nominal component values from effective parameters \\
Raw waveform & $V_\xi$, state voltages, time, run metadata & preserves the primary evidence before reduction \\
Reduction script & fixed zero-crossing and observable extraction rules & prevents result-dependent analysis changes \\
Evidence label & mathematical, ideal circuit, functional SPICE, vendor model, or physical hardware & prevents overstatement of certification level \\
\bottomrule
\end{tabular}
\end{table}

The eventual public package should therefore contain both human-readable documentation and machine-readable tables. Where a proprietary vendor macromodel cannot be redistributed, the package should retain the exact vendor source, version identifier, wrapper/netlist, and cryptographic hash so that the published result remains auditable without republishing restricted model files.

\subsection{Scope and limitations}

Several limitations remain. First, the calibrated-active regularized-$v$ vendor-macromodel simulation gate is closed, but vendor-level certification is not yet closed for the preferred common $w$-form architecture. Second, physical hardware evidence is absent, and that result does not establish installed ramp/offset uncertainty, leakage, loading, noise, aging, parasitics, or bench repeatability. Third, the self-closed radiative model uses deliberately simplified closures and is not a calibrated stellar model. Fourth, the literature search has been broadened across astronomy, engineering, historical analogue-computing, NASA/NTRS, and general scholarly indexes, but it is still not exhaustive enough to support a priority claim. These limitations are part of the project design and are not hidden by the high accuracy of the ideal mathematical tests or by the successful vendor-model engineering certification.

\section{Conclusions}

The completed mathematical and ideal scaled-voltage work establishes a reproducible pathway from static stellar equations to bounded-voltage analog-computing architectures. Exact and high-accuracy Lane--Emden references provide two global stellar observables, $\xi_1$ and $\mu_1$, against which circuit mappings can be tested. Central-series initialization removes the stellar-center singularity from the electronic representation, derivative-state conditioning avoids poorly measurable auxiliary signals, and computation speed can be adjusted to restore headroom without changing the target stellar solution. A common source-selectable platform can represent the $n=0$, $n=1$, and $n=3$ benchmark family with all audited ideal nodes below 9 V. Sensitivity and Monte Carlo studies further show that publication-grade comparison requires calibrated effective block parameters rather than nominal component values alone.

The evidence base extends beyond the mathematical and ideal scaled-voltage levels: the calibrated-active regularized-$v$ bridge passes its independently audited vendor-macromodel simulation engineering gate. That result is intentionally narrower than a 0.1\% scientific-accuracy certification and narrower than physical-hardware validation. Vendor execution of the preferred common $w$-form architecture and physical measurements therefore remain separate evidence levels. Taken together, the benchmark family, circuit scaling, calibrated uncertainty model, and independently audited vendor-macromodel bridge establish a reproducible computational-methods demonstration of electronic analog stellar-structure calculation without representing unbuilt hardware as experimental evidence.

\section*{Data, code, and materials availability}
The machine-readable result tables underlying the independently audited regularized-$v$ certification, the ideal benchmark scripts, circuit-source files, manuscript figure inputs, calibration specifications, and audit summaries are supplied in the accompanying reproducibility release candidate. The exact archive distributed with this manuscript is identified by a SHA-256 checksum in the submission package. A DOI-bearing archival copy should be deposited with the final public release. Official vendor macromodel files are not redistributed where their licenses do not permit redistribution; instead, the release records model provenance and the cryptographic identity of the simulation package. No physical-hardware waveform is claimed or supplied because physical validation is outside the evidence level of this paper.

\clearpage
\section*{Supplementary material}
Detailed archived exploratory prototype circuit sheets, the common-platform source-selector coefficient audit, calibration/run-control architecture, the extended literature-search protocol, and circuit-figure evidence-status table are provided in the accompanying \emph{Supplementary Methods and Circuit Documentation}. A separate \emph{Physical Circuit Atlas} records the design-stage physical translation of the certified regularized-$v$ topology; it is explicitly not physical-hardware evidence. The main manuscript retains the circuit figures required to understand and audit the scientific method.

\end{document}


\maketitle
\renewcommand{\thesection}{S\arabic{section}}
\renewcommand{\thesubsection}{S\arabic{section}.\arabic{subsection}}
\renewcommand{\thefigure}{S\arabic{figure}}
\renewcommand{\thetable}{S\arabic{table}}
\setcounter{section}{0}
\setcounter{figure}{0}
\setcounter{table}{0}

This supplement preserves detailed circuit provenance and reproducibility
material that is useful for audit but not required in the narrative flow of
the main article. Evidence levels are unchanged: archived Multisim sheets are
historical provenance; the common $w$-form circuit is verified at the
mathematical/ideal scaled-voltage level; the calibrated regularized-$v$ device-level
realization has vendor-macromodel simulation certification; physical hardware remains pending.

\section{Archived exploratory circuit sheets}
\label{app:p0sheets}

The figures in this appendix reproduce the exploratory Multisim sheets supplied at the beginning of the project. They are retained as historical provenance. They are not treated as the final certified circuit design: the main text documents the sign, coefficient, central-singularity, and conditioning issues found by audit.

\begin{figure}[H]
\centering
\begin{subfigure}[t]{0.48\textwidth}
\centering
\includegraphics[width=\linewidth]{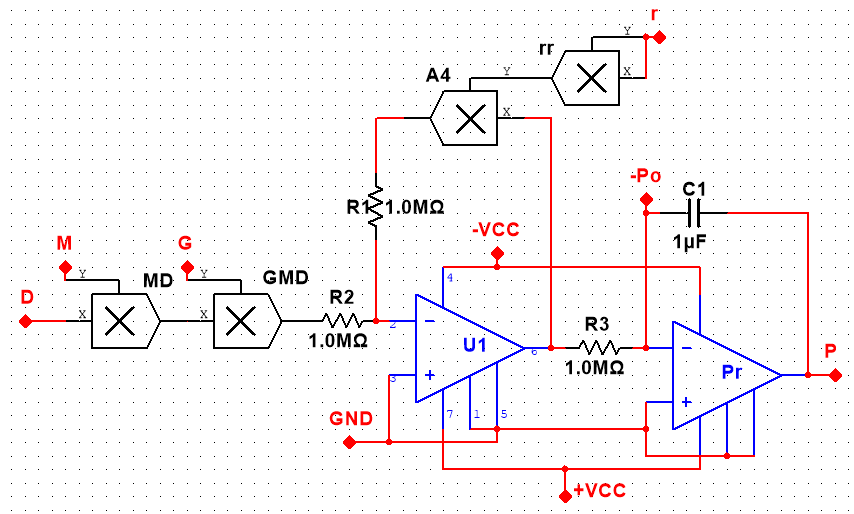}
\caption{Hydrostatic-equilibrium circuit sheet.}
\end{subfigure}\hfill
\begin{subfigure}[t]{0.48\textwidth}
\centering
\includegraphics[width=\linewidth]{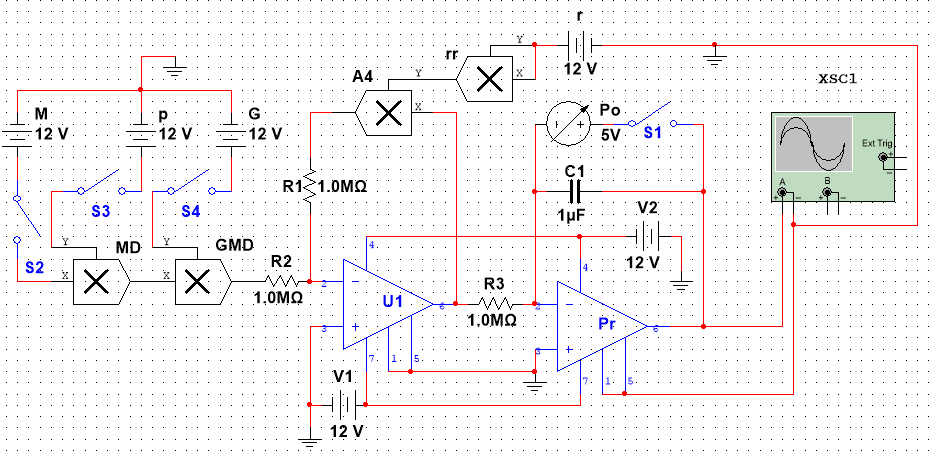}
\caption{Hydrostatic test/measurement sheet.}
\end{subfigure}
\caption{Exploratory prototype hydrostatic-equilibrium circuit material.}
\label{fig:app_hydrostatic}
\end{figure}

\begin{figure}[H]
\centering
\begin{subfigure}[t]{0.48\textwidth}
\centering
\includegraphics[width=\linewidth]{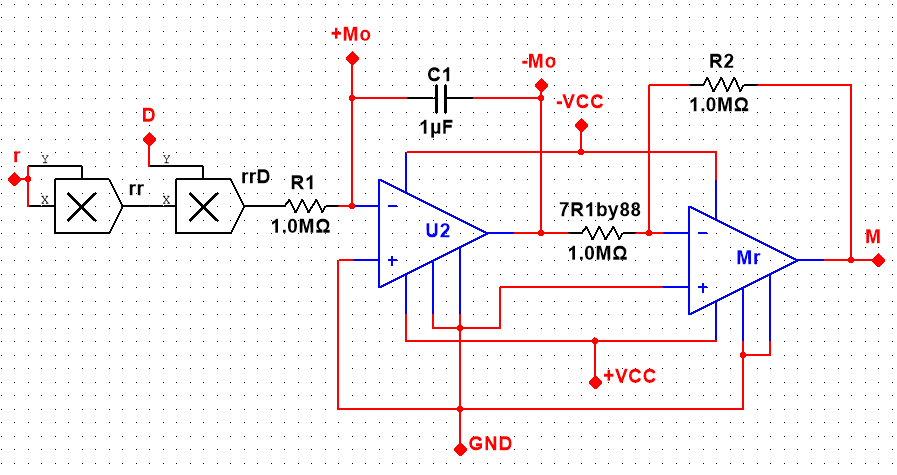}
\caption{Mass-continuity circuit sheet.}
\end{subfigure}\hfill
\begin{subfigure}[t]{0.48\textwidth}
\centering
\includegraphics[width=\linewidth]{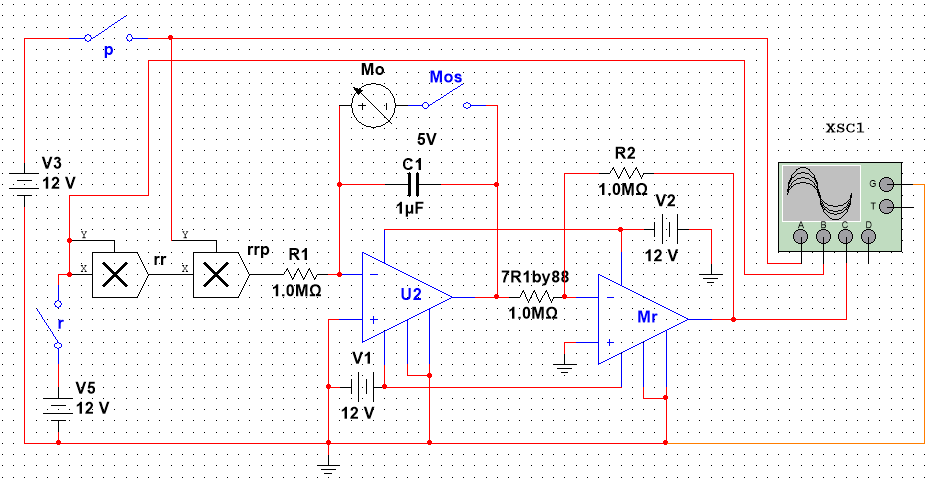}
\caption{Mass-continuity test/measurement sheet.}
\end{subfigure}
\caption{Exploratory prototype mass-continuity circuit material.}
\label{fig:app_mass}
\end{figure}

\begin{figure}[H]
\centering
\begin{subfigure}[t]{0.48\textwidth}
\centering
\includegraphics[width=\linewidth]{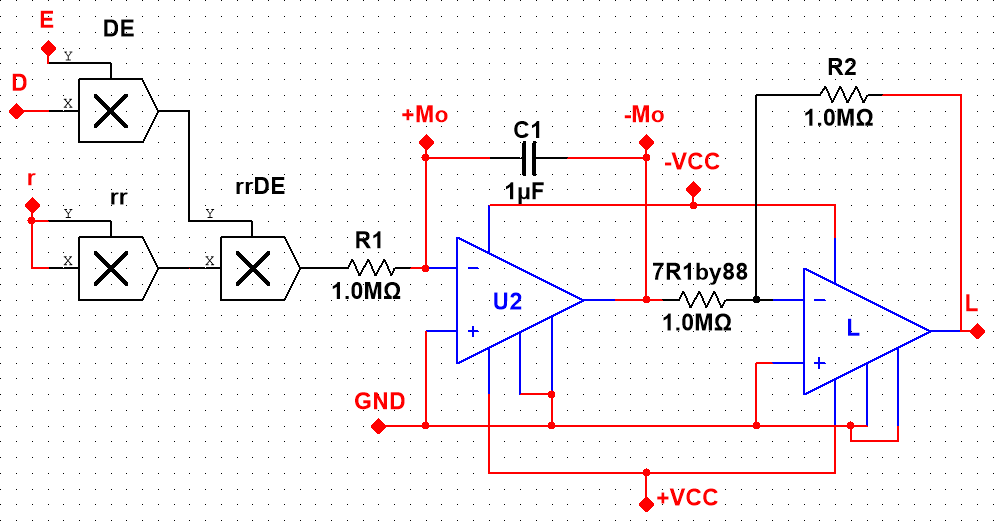}
\caption{Energy-generation circuit sheet.}
\end{subfigure}\hfill
\begin{subfigure}[t]{0.48\textwidth}
\centering
\includegraphics[width=\linewidth]{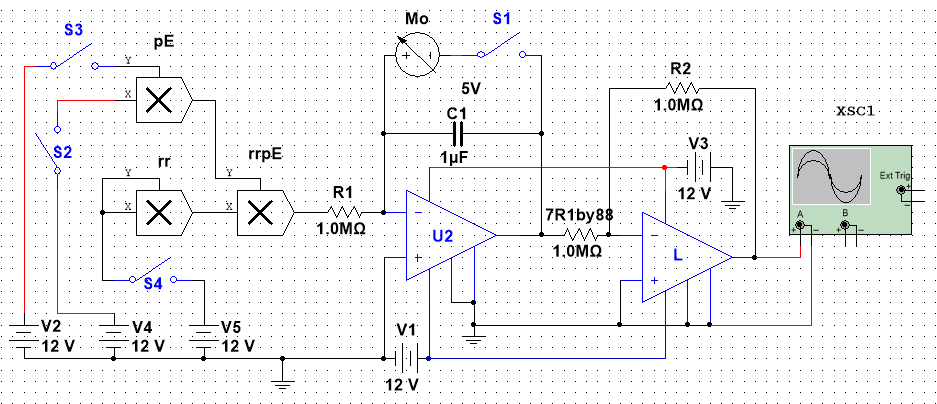}
\caption{Energy-generation test/measurement sheet.}
\end{subfigure}
\caption{Exploratory prototype energy-generation circuit material.}
\label{fig:app_energy}
\end{figure}

\begin{figure}[H]
\centering
\begin{subfigure}[t]{0.48\textwidth}
\centering
\includegraphics[width=\linewidth]{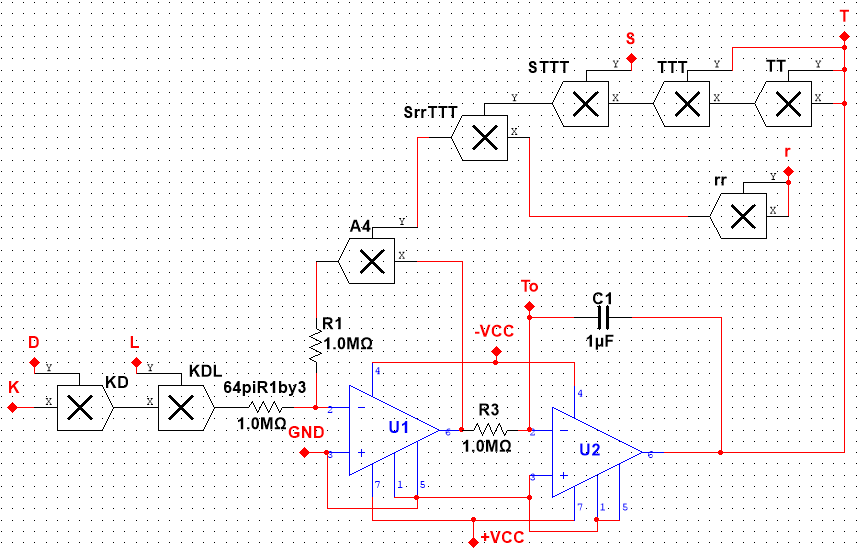}
\caption{Radiative-energy-transport circuit sheet.}
\end{subfigure}\hfill
\begin{subfigure}[t]{0.48\textwidth}
\centering
\includegraphics[width=\linewidth]{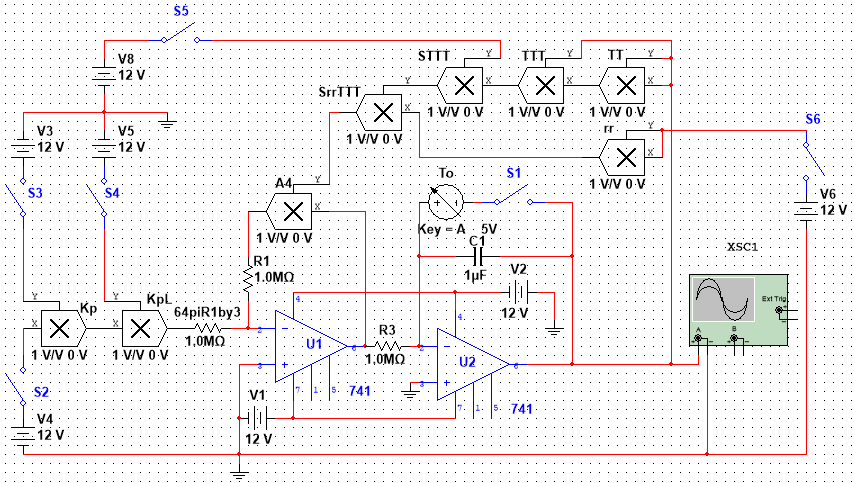}
\caption{Radiative test/measurement sheet.}
\end{subfigure}
\caption{Exploratory prototype radiative-transport circuit material.}
\label{fig:app_radiative}
\end{figure}

\begin{figure}[H]
\centering
\begin{subfigure}[t]{0.48\textwidth}
\centering
\includegraphics[width=\linewidth]{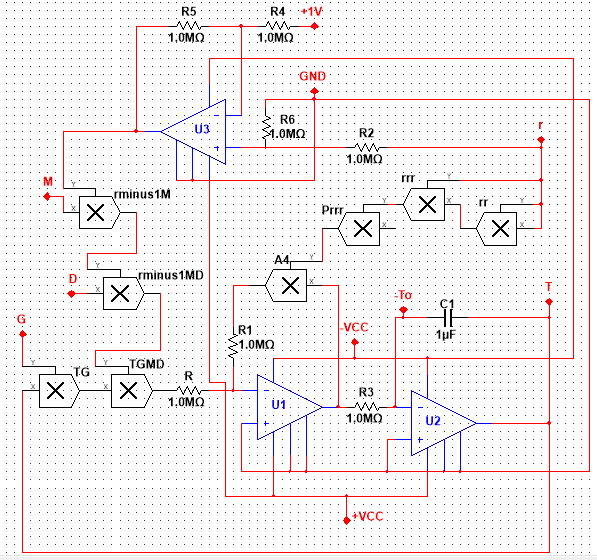}
\caption{Convective-energy-transport circuit sheet.}
\end{subfigure}\hfill
\begin{subfigure}[t]{0.48\textwidth}
\centering
\includegraphics[width=\linewidth]{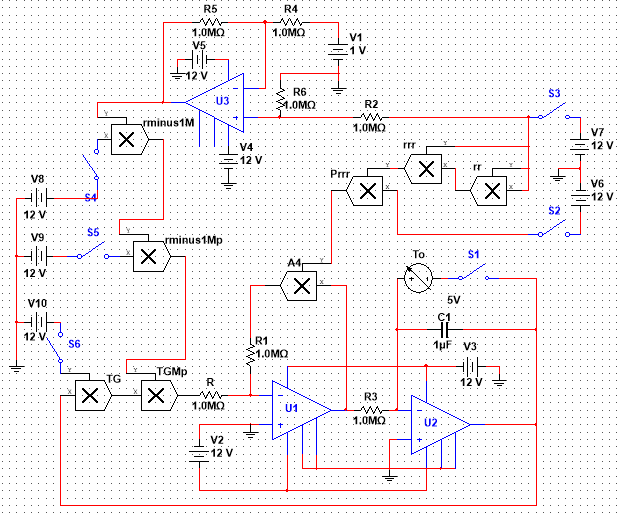}
\caption{Convective test/measurement sheet.}
\end{subfigure}
\caption{Exploratory prototype convective-transport circuit material.}
\label{fig:app_convective}
\end{figure}

\begin{figure}[H]
\centering
\includegraphics[width=0.82\textwidth]{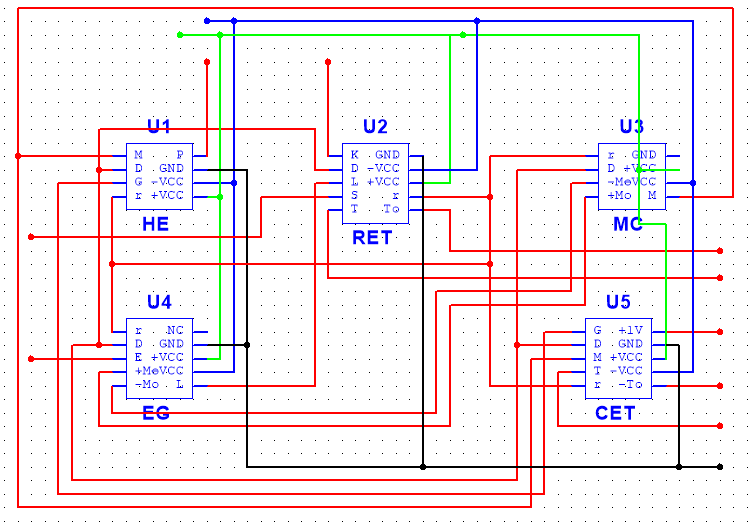}
\caption{Exploratory prototype combined five-equation wiring sheet. The density, opacity, energy-generation, mass, and transport subcircuits were interconnected before the later coefficient/sign/conditioning audit.}
\label{fig:app_combined}
\end{figure}

\clearpage
\section{Common-platform circuit details}
\label{app:commoncircuit}

\subsection{Polytropic source selector}
\label{app:sourceselector}

\begin{figure}[H]
\centering
\includegraphics[width=0.96\textwidth]{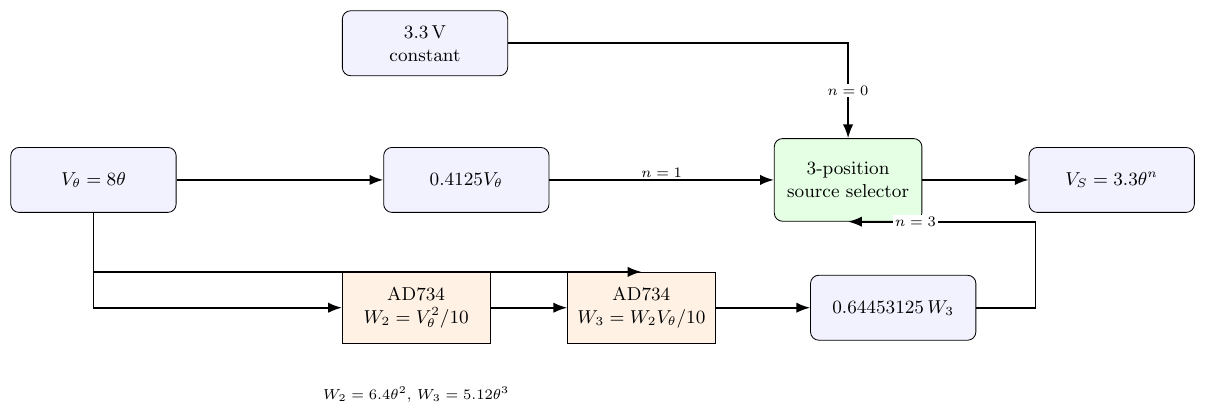}
\caption{Detailed source-selector branch for the common $n=0,1,3$ platform.}
\label{fig:app_selector}
\end{figure}

The common $w$-form backbone is unchanged across $n=0$, $n=1$, and $n=3$. The source branch alone produces $V_S=3.3\theta^n$ as shown in Figure~\ref{fig:app_selector}. For $n=3$, two AD734 multiplications give
\begin{equation}
W_2=\frac{V_\theta^2}{10}=6.4\theta^2,
\qquad
W_3=\frac{W_2V_\theta}{10}=5.12\theta^3,
\end{equation}
followed by the calibrated gain $0.64453125$.

\subsection{Coefficient audit of the common $w$-form}

With $V_\xi=\xi$, $V_\theta=8\theta$, $V_w=11w$, and $d\xi/d\tau=0.30$,
\begin{align}
\frac{dV_\theta}{d\tau}
&=8(0.30)(-w)
=-\frac{2.4}{11}V_w,\\
\frac{dV_w}{d\tau}
&=11(0.30)\left(\theta^n-\frac{2w}{\xi}\right)\\
&=3.3\theta^n-0.6\frac{V_w}{V_\xi}.
\end{align}
An inverting integrator therefore receives $+(2.4/11)V_w$ in the $\theta$ channel and $-V_S+0.6(V_w/V_\xi)$ in the $w$ channel. This sign accounting is the appendix-level circuit audit corresponding to the common-platform circuit figure in the main manuscript.

\section{Calibration and run-control architecture}
\label{app:control}

\begin{figure}[H]
\centering
\includegraphics[width=0.95\textwidth]{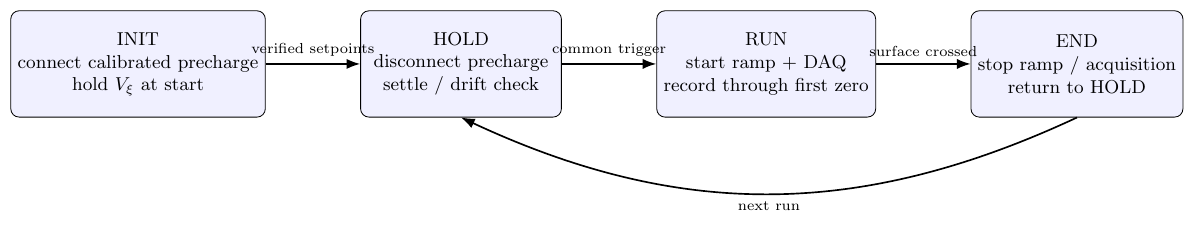}
\caption{Planned INIT/HOLD/RUN/END control sequence for calibrated precharge, switch settling, synchronized ramp start, and waveform acquisition.}
\label{fig:app_control}
\end{figure}

The future physical platform is intended to separate state initialization from computation. Figure~\ref{fig:app_control} summarizes the planned control states. The sequence is included as a reproducibility design, not as evidence of completed hardware.

\section{Literature-search protocol and current novelty position}
\label{app:literaturesearch}

The manuscript deliberately avoids a priority claim. Rosseland's Oslo Analyzer is now treated as an important institutional and astrophysical analogue-computing precedent: it was built at the University of Oslo Institute of Theoretical Astrophysics, and Holst documents its development and scientific role \citep{Holst1996}. Kopal later explicitly described Rosseland's work as a pioneer application of the differential analyzer to astronomical problems \citep{Kopal1949}. Analogue computation has
a genuine history in astrophysical problems: Lema\^itre and Vallarta used
Bush's mechanical differential analyzer to calculate cosmic-ray trajectories
in the Earth's dipolar magnetic field \citep{LemaitreVallarta1936}. More
importantly for the present subject, Kopal mechanically integrated the
nonradial-oscillation equations of the standard $n=3$ polytropic stellar model
on the Rockefeller differential analyzer at MIT \citep{Kopal1949}. Thus the
literature question is not whether analogue computation was ever used in
astronomy or even on polytropic stellar models, but whether a directly
comparable component-level \emph{electronic} realization of the \emph{static}
stellar-structure/Lane--Emden problem has been reported.

The September 2026 search combined historical analogue-computing literature,
astronomy and space archives, scholarly web indexes, and a semantic
academic-paper index. Five search families are especially important.

\begin{table}[htbp]
\centering
\small
\caption{Closest literature families identified in the current bounded search. The classifications describe the retrieved records and do not establish literature-wide priority.}
\label{tab:searchlog}
\begin{tabular}{p{0.21\textwidth}p{0.28\textwidth}p{0.40\textwidth}}
\toprule
Literature family & Representative precedents & Relevance and distinction \\
\midrule
Astrophysical analogue computation &
Lema\^itre \& Vallarta (1936) \citep{LemaitreVallarta1936} &
Bush differential-analyzer calculation of cosmic-ray trajectories; confirms genuine astrophysical analogue-computing precedent, but not stellar structure or Lane--Emden. \\
Stellar/polytropic analogue computation &
Kopal (1949) \citep{Kopal1949} &
Rockefeller differential-analyzer integration of nonradial oscillation equations for the standard $n=3$ polytropic stellar model; the machine used tabulated Emden background functions and a finite-radius series start. Direct stellar analogue-computing precedent, but a mechanical oscillation calculation rather than a static Lane--Emden electronic circuit. \\
Electronic differential analyzers &
Howe \& Haneman (1953) \citep{HoweHaneman1953}; Nagaraja et al. (1957) \citep{Nagaraja1957} &
Component-level feedback-amplifier analogue machines for linear, nonlinear, and spatially discretized differential equations; direct methodological precedent, but no stellar-structure implementation in the retrieved records. \\
Digital/numerical stellar structure &
Henyey et al. (1959) \citep{Henyey1959}; Hoyle (1959) \citep{Hoyle1959}; Larson \& Demarque (1964) \citep{LarsonDemarque1964} &
Direct stellar-evolution or stellar-structure computation, but using digital machines or numerical boundary-value methods rather than a component-level continuous-voltage analogue realization. \\
Lane--Emden numerical methods &
Representative modern numerical and operational-matrix treatments \citep{Kumar2011,Anandaram2013} &
Direct benchmark-domain comparators; useful for reference solutions, but not electronic analogue hardware. \\
\bottomrule
\end{tabular}
\end{table}

One historical stellar-interior paper requires special caution. Masani and
Occheni discussed disturbances and shock waves inside stars and their
solubility by ``electronic computers'' \citep{MasaniOccheni1962}. The indexed
abstract does not establish an analogue-computer circuit, so the work is not
classified here as analogue evidence without full-text confirmation. The same
discipline is applied throughout this review: the words \emph{machine} or
\emph{electronic computer} are not silently reinterpreted as \emph{analogue
computer}.

Across the searches completed to date, no retrieved record has been a direct
match to the complete combination used here for the \emph{static}
Lane--Emden problem: continuous-voltage stellar state mapping, component-level
\emph{electronic} integrators and nonlinear blocks, central regularization,
multiple polytropic benchmarks, global observables $\xi_1$ and $\mu_1$,
bounded-voltage design, and calibrated uncertainty propagation. Kopal's
mechanical stellar-oscillation computation is therefore treated as a close
historical analogue-computing precedent, not as the same implementation class.
The remaining absence is a negative search result rather than proof that no
prior implementation exists.

The submission-stage wording is therefore intentionally bounded: this work
\emph{develops and verifies the present circuit architecture}. It does not
claim to be the first analogue computer used in astrophysics, the first
analogue computation applied to a stellar model, the first electronic solution
of the Lane--Emden equation, or the first electronic stellar model. Before any narrower priority language could be considered, the
historical search should be completed with direct full-text verification,
backward/forward citation chaining, and searches in the major astronomy,
engineering, and library indexes.

\section{Circuit-figure status and claim limits}

Table~\ref{tab:schematicstatus} prevents the circuit figures from being interpreted as evidence beyond their actual certification level.

\begin{table}[H]
\centering
\caption{Status of circuit figures and design documents accompanying the manuscript.}
\label{tab:schematicstatus}
\begin{tabular}{p{0.27\textwidth}p{0.25\textwidth}p{0.38\textwidth}}
\toprule
Circuit figure & Evidence level & Current claim limit \\
\midrule
Exploratory Multisim sheets & historical exploratory design & provenance only; not certified \\
Cleaned hydrostatic map & audited mathematical/signal-flow reconstruction & sign/operation mapping only \\
Common $w$-form architecture & ideal scaled-voltage architecture & mathematical and ideal-circuit verification complete; preferred vendor execution pending \\
Source-selector detail & ideal nonlinear-block architecture & coefficient mapping complete; installed-device calibration pending \\
Regularized-$v$ $n=1$ device-level bridge & vendor-macromodel simulation certification & predefined calibrated-active engineering gate passed after independent audit; physical implementation pending \\
INIT/HOLD/RUN control & future physical design & planning only; no hardware evidence \\
\bottomrule
\end{tabular}
\end{table}

\section{Design-stage physical translation of the certified regularized-$v$ topology}

The publication package also includes the separate file \emph{Physical Circuit Atlas}. It translates the simulation-certified regularized-$v$ topology into a capture-stage physical circuit design with explicit multiplier/divider paths, resistor ratios, state integrators, reset and null-current networks, candidate low-leakage relay control, external reference interfaces, calibration test points, and PCB partitioning. This atlas is included to make the circuit architecture inspectable and to document how the simulation topology could be realized physically. It is \emph{not} experimental evidence: no board-level waveform, installed leakage, calibration stability, or physical $\xi_1/\mu_1$ result is claimed in the present paper.